\documentclass[12pt]{article}
\usepackage{amssymb}
\usepackage{epsf}
\usepackage{epsfig}
\newcommand{\lsim}{
\mathrel{\hbox{\rlap{\hbox{\lower4pt\hbox{$\sim$}}}\hbox{$<$}}}}

\newcommand{\gsim}{
\mathrel{\hbox{\rlap{\hbox{\lower4pt\hbox{$\sim$}}}\hbox{$>$}}}}

\newcommand{\be}{\begin{equation}}
\newcommand{\ee}{\end{equation}}
\newcommand{\bi}{\begin{itemize}}
\newcommand{\ei}{\end{itemize}}

\begin{document}
\begin{titlepage}
\vspace*{-0.5truecm}

\begin{flushright}
CERN-PH-TH/2004-216\\
TUM-HEP-561/04\\
MPP-2004-141\\
hep-ph/0410407
\end{flushright}

\vspace*{0.3truecm}

\begin{center}
\boldmath
{\Large{\bf The $B\to\pi\pi,\pi K$ Puzzles in the Light of New Data: 

\vspace{0.3truecm}

Implications for the Standard Model, 

\vspace{0.4truecm}

New Physics and Rare Decays}}
\unboldmath
\end{center}

\vspace{0.5truecm}

\begin{center}
{\bf Andrzej J. Buras,${}^a$ Robert Fleischer,${}^b$ 
Stefan Recksiegel${}^a$ and Felix Schwab${}^{c,a}$}
 
\vspace{0.4truecm}

${}^a$ {\sl Physik Department, Technische Universit\"at M\"unchen,
D-85748 Garching, Germany}

\vspace{0.2truecm}

${}^b$ {\sl Theory Division, Department of Physics, CERN, 
CH-1211 Geneva 23, Switzerland}

\vspace{0.2truecm}

 ${}^c$ {\sl Max-Planck-Institut f{\"u}r Physik -- Werner-Heisenberg-Institut,
 D-80805 Munich, Germany}

\end{center}

\vspace{0.7cm}
\begin{abstract}
\vspace{0.2cm}\noindent
Recently, we developed a strategy to analyse the $B\to\pi\pi,\pi K$ data. 
We found that the $B\to\pi\pi$ measurements can be accommodated in 
the Standard Model (SM) through large non-factorizable effects. On 
the other hand, our analysis of the ratios $R_{\rm c}$ and 
$R_{\rm n}$ of the CP-averaged branching ratios of the charged and neutral 
$B\to\pi K$ modes, respectively, suggested new physics (NP) in the 
electroweak penguin sector, which may have a powerful interplay with rare 
decays. In this paper, we confront our strategy with recent experimental 
developments, addressing also the direct CP violation in $B_d\to\pi^\mp K^\pm$,
which is now an established effect, the relation to its counterpart in 
$B^\pm\to\pi^0K^\pm$, and the first results for the direct CP asymmetry of 
$B_d\to\pi^0\pi^0$ that turn out to be in agreement with our prediction. We 
obtain hadronic $B\to\pi\pi,\pi K$ parameters which are almost unchanged and 
arrive at an allowed region for the unitarity triangle in perfect accordance 
with the SM. The ``$B\to\pi K$ puzzle'' persists, and can still be explained 
through NP, as in our previous analysis. In fact, the recently observed shifts 
in the experimental values of $R_{\rm c}$ and $R_{\rm n}$ have been predicted 
in our framework on the basis of constraints from rare decays. Conversely, 
we obtain a moderate deviation of the ratio $R$ of the CP-averaged 
$B_d\to\pi^\mp K^\pm$ and $B^\pm\to\pi^\pm K$ rates from the current 
experimental value. However, using the emerging signals for $B^\pm\to K^\pm K$ 
modes, this effect can be attributed to certain hadronic effects, which have 
a minor impact on $R_{\rm c}$ and do not at all affect $R_{\rm n}$. Our 
results for rare decays remain unchanged.
\end{abstract}

\vspace*{0.5truecm}
\vfill
\noindent
October 2004

\end{titlepage}

\thispagestyle{empty}
\vbox{}
\newpage

\setcounter{page}{1}

\section{Introduction}\label{sec:intro}
\setcounter{equation}{0}
A particularly interesting aspect of the physics programme of the $B$ 
factories for the exploration of the Kobayashi--Maskawa mechanism of CP 
violation \cite{KM} is given by $B\to\pi\pi$ and $B\to\pi K$ modes 
(see \cite{BaBar-Book,RF-Phys-Rep} and references therein). In the 
summer of 2003, the last missing element of the $B\to\pi\pi$ system, 
$B^0_d\to \pi^0\pi^0$, was eventually observed by the 
BaBar~\cite{Babar-Bpi0pi0} and Belle collaborations \cite{Belle-Bpi0pi0}, 
with a surprisingly prominent rate, pointing to large corrections to the 
predictions of the QCD factorization approach \cite{BBNS1,Be-Ne}. 
Concerning the $B\to\pi K$ system, the last missing transition, 
$B^0_d\to \pi^0 K^0$, was already observed by the CLEO collaboration 
in 2000 \cite{CLEO-BpiK}, and is now well established. Immediately after 
that measurement, a puzzling pattern in the ratios 
$R_{\rm c}$ and $R_{\rm n}$ of the charged and neutral $B\to\pi K$ 
rates has been pointed out \cite{BF-neutral2}. This ``$B\to\pi K$ puzzle'' 
has survived over the years and was recently reconsidered by several authors
(see, for instance, \cite{Be-Ne,yoshikawa,gr-ewp,BFRS-I}). 
It is an important feature of the observables $R_{\rm c}$ and $R_{\rm n}$ 
that they are significantly affected by (colour-allowed) electroweak (EW) 
penguins \cite{BF-neutral1,neubert}. On the other hand, the ratio $R$ 
of the CP-averaged $B_d\to\pi^\mp K^\pm$ and $B^\pm\to\pi^\pm K$ branching 
ratios, which is expected to be only marginally affected by 
(colour-suppressed) EW penguins, does not show an anomalous behaviour. 
Since EW penguins offer an attractive avenue for new physics (NP) to 
manifest itself \cite{FM-NP,trojan}, the $B\to\pi K$ puzzle may indicate 
NP in the EW penguin sector. Should this actually be the case, also 
several rare $K$ and $B$ decays may show NP effects \cite{BFRS-I}, thereby 
complementing the $B\to\pi K$ puzzle in a valuable manner.

In \cite{BFRS-PRL,BFRS-BIG}, we developed a strategy to address these 
exciting issues systematically, with the following logical structure:
\begin{itemize}
\item[i)]Using the isospin flavour symmetry of strong interactions and 
assuming the range for the angle $\gamma$ of the unitarity triangle (UT) 
that follows in the Standard Model (SM) from the CKM fits
\cite{CKM-Book, CKM-fitters, Bona:2004sj}, 
we may extract a  set of hadronic parameters characterizing the 
$B\to\pi\pi$ system from the 
experimental results for the corresponding CP-averaged branching ratios 
and the CP-violating observables of $B_d\to\pi^+\pi^-$. We found 
large deviations of the hadronic parameters from the predictions of 
QCD factorization. Moreover, we could predict the CP asymmetries of the 
$B_d\to\pi^0\pi^0$ decay in the SM. 
\item[ii)]If we use the $SU(3)$ flavour symmetry and neglect penguin
annihilation and exchange topologies, which can be probed through 
$B_d\to K^+K^-$ and $B_s\to\pi^+\pi^-$ modes, the hadronic $B\to\pi\pi$ 
parameters allow us to determine their $B\to\pi K$ counterparts. Assuming
again the SM, as in the $B\to\pi\pi$ analysis, we may predict all observables
offered by the $B\to\pi K$ system, including also CP asymmetries. We found 
agreement with the experimental picture for $R$, whereas the situation 
in the $R_{\rm n}$--$R_{\rm c}$ plane was {\it not} in accordance with 
experiment. This discrepancy could be resolved through NP effects in the 
EW penguin sector, requiring a significant enhancement of the parameter 
$q$ measuring their strength relative to the tree contributions, and a 
NP phase $\phi$, which vanishes in the SM, around $-90^\circ$.
\item [iii)]Assuming a more specific (but popular) scenario
\cite{Colangelo:1998pm}--\cite{Buchalla:2000sk}, where NP enters 
the EW penguin sector through $Z^0$ penguins \footnote{See
 \cite{Barger:2004hn} for a discussion of the $B \to \pi K$ system in a
 slightly different scenario involving an additional $Z^{'}$ boson.}, 
we obtain an interesting 
interplay between the parameters $q$ and $\phi$ following from the resolution 
of the $B\to\pi K$ puzzle and several rare $B$ and $K$ decays. This allowed 
us to explore the impact of the data for $B\to X_s\mu^+\mu^-$ and 
$K_{\rm L}\to\pi^0e^+e^-$ processes, constraining the enhancement of
$q$ to be smaller than that suggested by the $B\to\pi K$ data, thereby favouring 
a smaller value of $R_{\rm c}$ and a larger value of $R_{\rm n}$. In 
fact, this pattern has been confirmed to a large extent by the new data.
Taking these constraints into account, there may still be prominent 
NP effects in the rare-decay sector, the most spectacular ones in 
$K_{\rm L}\to\pi^0\nu\bar\nu$ and $B_{s,d}\to\mu^+\mu^-$, exhibiting 
branching ratios that could be enhanced with respect to the SM by 
factors of ${\cal O}(10)$ and ${\cal O}(5)$, respectively. 
\end{itemize}
In addition, we discussed the determination of $\gamma$ (and the other 
two UT angles $\alpha$ and $\beta$), where we obtained a result 
in agreement with the CKM fits \cite{CKM-Book, CKM-fitters, Bona:2004sj}, had a closer look at the 
$B_s$-meson decays $B_s\to K^+K^-$ and $B_s\to \pi^\pm K^\mp$, 
and performed a couple of consistency checks of the $SU(3)$ flavour 
symmetry, which did not indicate large corrections. 

As there were several exciting experimental developments thanks to 
the BaBar and Belle collaborations since we wrote our original papers 
\cite{BFRS-PRL,BFRS-BIG}, it is interesting to confront our strategy 
with the most recent data, although the picture is still far from 
being settled. The most important aspects are the following:
\begin{itemize}
\item Several new results for the $B\to\pi\pi$ and $B\to\pi K$ branching
ratios \cite{BaBar-BK0pi0}--\cite{Belle-Bpi0pi0-new}.
\item First results for the direct CP asymmetry of 
$B_d\to\pi^0\pi^0$ \cite{BaBar-Bpi0pi0,Belle-Bpi0pi0-new}. 
\item Updates for the CP-violating observables of $B_d\to\pi^+\pi^-$
\cite{BaBar-CP-Bpipi,Belle-CP-Bpipi}, as well as for the CP asymmetries of 
several $B\to\pi K$ modes 
\cite{BaBar-BK0pi0,BaBar-Bpi0pi0,Belle-BpiK-CP,Belle-BK0pi0}. 
\item Observation of direct CP violation in $B_d\to\pi^\mp K^\pm$
\cite{BaBar-CP-dir-obs,Belle-CP-dir-obs}, representing a new milestone
in the exploration of CP violation. Some implications of this
measurement have been discussed in \cite{Wu:2004xx} and \cite{He:2004wh},
concerning the SM and supersymmetry, respectively.
\item Observation of $B_d\to K^0\bar K^0$, i.e.\ of the first $b\to d$ 
penguin decay, and an emerging signal for its charged counterpart 
$B^\pm\to K^\pm K$ \cite{BaBar-BdKK-obs}. 
\end{itemize}
In our analysis, we will use the averages for these new results that were
compiled by the ``Heavy Flavour Averaging Group'' (HFAG) \cite{HFAG},
but also make a number of refinements and generalizations, in particular:
\begin{itemize}
\item We include the EW penguins of the $B \to \pi \pi$ system in our 
analysis. As anticipated in \cite{BFRS-BIG}, this has a small impact 
on the numerics, but is a conceptual improvement.
\item In view of the new $B \to \pi K$ data, we investigate the impact 
of certain hadronic effects, which can be constrained through the 
emerging experimental signal for $B^\pm\to K^\pm K$ decays. 
\end{itemize}
The outline is as follows: in Section~\ref{sec:Bpipi}, we focus on the 
$B\to\pi\pi$ system and move on to the $B\to\pi K$ modes in 
Section~\ref{sec:BpiK}. Finally, we summarize our conclusions in 
Section~\ref{sec:concl}.

\boldmath
\section{The $B\to\pi\pi$ System}\label{sec:Bpipi}
\unboldmath
\setcounter{equation}{0}
\subsection{Amplitudes}\label{ssec:Bpipi-ampls}
The starting point of our analysis of the $B\to\pi\pi$ decays is given
by the amplitudes
\begin{eqnarray}
\sqrt{2}A(B^+\to\pi^+\pi^0)&=&-[\tilde T+\tilde C] = 
-[T+C]\label{B+pi+pi0}\\
A(B^0_d\to\pi^+\pi^-)&=&-[\tilde T + P]\label{Bdpi+pi-}\\
\sqrt{2}A(B^0_d\to\pi^0\pi^0)&=&-[\tilde C - P],\label{Bdpi0pi0}
\end{eqnarray}
which satisfy the following well-known isospin relation \cite{grolo}:
\begin{equation}\label{Bpipi-isospin}
\sqrt{2}A(B^+\to\pi^+\pi^0)=A(B^0_d\to\pi^+\pi^-)+
\sqrt{2}A(B^0_d\to\pi^0\pi^0).
\end{equation}
The individual amplitudes of (\ref{B+pi+pi0})--(\ref{Bdpi0pi0}) can
be expressed as
\begin{eqnarray}
P&=&\lambda^3 A({\cal P}_t-{\cal P}_c)\equiv\lambda^3 A {\cal P}_{tc}
\label{P-def}\\
\tilde T &=&\lambda^3 A R_b e^{i\gamma}\left[{\cal T}-\left({\cal 
P}_{tu}-{\cal E}\right)\right]\label{T-tilde}\\
\tilde C &=&\lambda^3 A R_b e^{i\gamma}\left[{\cal C}+\left({\cal P}_{tu}-
{\cal E}\right)\right],\label{C-tilde}
\end{eqnarray}
where 
\begin{equation}
\lambda\equiv|V_{us}|=0.2240\pm 0.0036, \quad
A\equiv |V_{cb}|/\lambda^2=0.83\pm0.02 
\end{equation}
are the usual parameters in the Wolfenstein expansion of the 
Cabibbo--Kobayashi--Maskawa (CKM) matrix \cite{WO,BLO},
\begin{equation}\label{Rb-def}
R_b\equiv \sqrt{\bar\rho^2 + \bar\eta^2}=
\left(1-\frac{\lambda^2}{2}\right)
\frac{1}{\lambda}\left|\frac{V_{ub}}{V_{cb}}\right|=
0.37\pm0.04
\end{equation}
measures one side of the UT, the ${\cal P}_q$ describe the strong 
amplitudes of QCD penguins with internal $q$-quark exchanges 
($q\in\{t,c,u\}$), including annihilation and exchange penguins,
while ${\cal T}$ and ${\cal C}$ are the strong amplitudes of 
colour-allowed and colour-suppressed tree-diagram-like topologies, 
respectively, and ${\cal E}$ denotes the strong amplitude of an 
exchange topology. The amplitudes $\tilde T$ and $\tilde C$ differ from
\begin{equation}
T =\lambda^3 A R_b e^{i\gamma}{\cal T}, \quad
C =\lambda^3 A R_b e^{i\gamma}{\cal C}
\end{equation}
through the $({\cal P}_{tu}-{\cal E})$ pieces, which may play
an important r\^ole \cite{PAP0}. Note that these terms contain 
also the ``GIM penguins'' with internal up-quark exchanges, 
whereas their ``charming penguin'' counterparts enter in $P$ 
through ${\cal P}_{c}$, as can be seen in (\ref{P-def}) 
\cite{PAP0}--\cite{BPRS}. In order to characterize the dynamics of the 
$B\to\pi\pi$ system, it is convenient to introduce the following 
hadronic parameters:
\begin{equation}\label{x-Bpipi}
x e^{i\Delta}\equiv \frac{\tilde C}{\tilde T}=
\left|\frac{\tilde C}{\tilde T}\right|e^{i(\delta_{\tilde C}-
\delta_{\tilde T})}=\frac{{\cal C}+\left({\cal P}_{tu}-
{\cal E}\right)}{{\cal T}-\left({\cal 
P}_{tu}-{\cal E}\right)}
\end{equation}
\begin{equation}\label{d-theta-def}
d e^{i\theta}\equiv-\frac{P}{\tilde T}e^{i\gamma}=
-\left|\frac{P}{\tilde T}\right|e^{i(\delta_P-\delta_{\tilde T})}
=-\frac{1}{R_b}\left[\frac{{\cal P}_{tc}}{{\cal T}-\left({\cal 
P}_{tu}-{\cal E}\right)}\right],
\end{equation}
where $\delta_{\tilde C}$, $\delta_{\tilde T}$ and $\delta_P$ denote 
the CP-conserving strong phases of $\tilde C$, $\tilde T$ and $P$.

In the $B\to\pi\pi$ system, the EW penguin contributions are expected
to play a minor r\^ole \cite{GHLR-EWP,PAPIII}, and were therefore 
neglected in (\ref{B+pi+pi0})--(\ref{Bdpi0pi0}). However, applying 
the isospin flavour symmetry of strong interactions, they can be 
included \cite{BF-neutral1,GPY}, yielding 
\begin{eqnarray}
\sqrt{2}A(B^+\to\pi^+\pi^0)&=&-|\tilde T|e^{i\delta_{\tilde T}}
\left[1+x e^{i\Delta}\right]\left[e^{i\gamma}+\tilde qe^{-i\beta}\right]
\label{B+pi+pi0EWP}\\
A(B^0_d\to\pi^+\pi^-)&=&-|\tilde T|e^{i\delta_{\tilde T}}
\left[e^{i\gamma}-d e^{i\theta}\right]\label{Bdpi+pi-EWP}\\
\sqrt{2}A(B^0_d\to\pi^0\pi^0)&=&|P|e^{i\delta_P} 
\left[1+\frac{x}{d}e^{i\gamma}e^{i(\Delta-\theta)}+\tilde q\left(
\frac{1+x e^{i\Delta}}{d}\right)e^{-i\theta}e^{-i\beta}
\right],\label{Bdpi0pi0EWP}
\end{eqnarray}
where
\begin{equation}\label{qtilde}
\tilde q\equiv\left|\frac{P_{\rm EW}}{T+C}\right|\approx1.3\times10^{-2}\times
\left|\frac{V_{td}}{V_{ub}}\right|=
1.3\times10^{-2}\times\left(1-\frac{\lambda^2}{2}\right)
\left|\frac{\sin\gamma}{\sin\beta}\right|\approx 3\times10^{-2}
\end{equation}
measures the strength of the sum of the colour-allowed and colour-suppressed
EW penguin contributions with respect to the sum of the colour-allowed 
and colour-suppressed tree-diagram-like contributions. It should be
emphasized that (\ref{qtilde}) was derived for the SM. In contrast to
our previous analysis \cite{BFRS-PRL,BFRS-BIG}, we shall also include
the EW penguin contributions in the numerical analysis performed below. 
Although their impact is actually small, this is a conceptual 
improvement. However, as soon as we consider NP in the EW penguin sector, 
the $B\to\pi\pi$ analysis does no longer fully separate from that of the 
$B\to\pi K$ system, i.e.\ items i) and ii) of Section~\ref{sec:intro}
are no longer completely independent. However, their cross talk is
actually very small. 

\subsection{Input Observables}
Following \cite{BFRS-PRL,BFRS-BIG}, we use the ratios
\begin{eqnarray}
R_{+-}^{\pi\pi}&\equiv&2\left[\frac{\mbox{BR}(B^+\to\pi^+\pi^0)
+\mbox{BR}(B^-\to\pi^-\pi^0)}{\mbox{BR}(B_d^0\to\pi^+\pi^-)
+\mbox{BR}(\bar B_d^0\to\pi^+\pi^-)}\right]
\frac{\tau_{B^0_d}}{\tau_{B^+}}\label{Rpm-def}\\
R_{00}^{\pi\pi}&\equiv&2\left[\frac{\mbox{BR}(B_d^0\to\pi^0\pi^0)+
\mbox{BR}(\bar B_d^0\to\pi^0\pi^0)}{\mbox{BR}(B_d^0\to\pi^+\pi^-)+
\mbox{BR}(\bar B_d^0\to\pi^+\pi^-)}\right]\label{R00-def}
\end{eqnarray}
of the CP-averaged $B\to\pi\pi$ branching ratios, and the CP-violating 
observables provided by the time-dependent rate asymmetry 
\begin{eqnarray}
\lefteqn{\frac{\Gamma(B^0_d(t)\to \pi^+\pi^-)-\Gamma(\bar B^0_d(t)\to 
\pi^+\pi^-)}{\Gamma(B^0_d(t)\to \pi^+\pi^-)+\Gamma(\bar B^0_d(t)\to 
\pi^+\pi^-)}}\nonumber\\
&&={\cal A}_{\rm CP}^{\rm dir}(B_d\to \pi^+\pi^-)\cos(\Delta M_d t)+
{\cal A}_{\rm CP}^{\rm mix}(B_d\to \pi^+\pi^-)
\sin(\Delta M_d t)\label{rate-asym}
\end{eqnarray}
as the input for our $B\to\pi\pi$ analysis. Concerning the former
quantities, they can be written in the following generic form:
\begin{equation}\label{Rpipi-gen}
R_{+-}^{\pi\pi}=F_1(d,\theta,x,\Delta;\gamma), \quad
R_{00}^{\pi\pi}=F_2(d,\theta,x,\Delta;\gamma).
\end{equation}
On the other hand, the CP-violating $B_d\to\pi^+\pi^-$ observables involve,
in addition to the angle $\gamma$ of the UT, only the hadronic parameters 
$(d,\theta)$; the mixing-induced CP asymmetry depends, furthermore, on the 
$B^0_d$--$\bar B^0_d$ mixing phase $\phi_d$, which equals $2\beta$
in the SM. Consequently, we may write
\begin{equation}\label{CP-Bpipi-gen}
{\cal A}_{\rm CP}^{\rm dir}(B_d\to \pi^+\pi^-)=
G_1(d,\theta;\gamma), \quad
{\cal A}_{\rm CP}^{\rm mix}(B_d\to \pi^+\pi^-)=
G_2(d,\theta;\gamma,\phi_d).
\end{equation}
Explicit expressions for the functions $F_{1,2}$ and $G_{1,2}$ can 
be found in \cite{BFRS-BIG}.

\begin{table}
\vspace{0.4cm}
\begin{center}
\begin{tabular}{|c||c|c|}
\hline
Quantity & This work & Previous analysis 
\\ \hline
 \hline
$\mbox{BR}(B^\pm\to\pi^\pm\pi^0)/10^{-6}$ & $5.5\pm0.6$ & $5.2 \pm 0.8$
\\ \hline
$\mbox{BR}(B_d\to\pi^+\pi^-)/10^{-6}$ & $4.6\pm0.4$ & $4.6\pm 0.4$
\\ \hline
$\mbox{BR}(B_d\to\pi^0\pi^0)/10^{-6}$ & $1.51\pm0.28$ & $1.9 \pm 0.5$
\\ \hline
 \hline
$R_{+-}^{\pi\pi}$ & $2.20\pm0.31$ & $2.12\pm0.37$
\\ \hline
$R_{00}^{\pi\pi}$ & $0.67\pm0.14$ & $0.83\pm0.23$
\\ \hline
 \hline
${\cal A}_{\rm CP}^{\rm dir}(B_d\to \pi^+\pi^-)$ & $-0.37\pm 0.11$ & 
$-0.38\pm0.16$
\\ \hline
${\cal A}_{\rm CP}^{\rm mix}(B_d\to \pi^+\pi^-)$ & $+0.61\pm0.14$ & 
$+0.58\pm0.20$
\\ \hline
\end{tabular}
\end{center}
\caption[]{The current status of the $B\to\pi\pi$ input data for our
strategy, with averages taken from \cite{HFAG}, and comparison with the 
picture of our previous analysis \cite{BFRS-PRL,BFRS-BIG}. For the 
evaluation of $R_{+-}^{\pi\pi}$, we have used the life-time ratio
$\tau_{B^+}/\tau_{B^0_d}=1.086\pm0.017$ \cite{PDG}.}\label{tab:Bpipi-input}
\end{table}

In Table~\ref{tab:Bpipi-input}, we have summarized the current 
experimental situation of the $B\to\pi\pi$ observables that serve
as an input for our strategy, comparing also with the values that
we used for our previous analysis. Concerning the CP-averaged 
$B\to\pi\pi$ branching ratios, the values obtained by 
the BaBar \cite{BaBar-Bpi0pi0} and Belle \cite{Belle-Bpi0pi0-new} 
collaborations are in accordance with one another. On the other hand, 
the picture of the CP-violating $B_d\to\pi^+\pi^-$ observables is still 
not yet settled. The BaBar and Belle results are now given as follows:
\begin{equation}\label{Adir-exp-new}
{\cal A}_{\rm CP}^{\rm dir}(B_d\to\pi^+\pi^-)
=\left\{
\begin{array}{ll}
-0.09 \pm 0.15 \pm 0.04 & \mbox{(BaBar \cite{BaBar-CP-Bpipi}),}\\
-0.58 \pm 0.15 \pm 0.07 & \mbox{(Belle \cite{Belle-CP-Bpipi}),}
\end{array}
\right.
\end{equation}
\begin{equation}\label{Amix-exp-new}
{\cal A}_{\rm CP}^{\rm mix}(B_d\to\pi^+\pi^-)
=\left\{
\begin{array}{ll}
+0.30 \pm 0.17 \pm 0.03 & \mbox{(BaBar \cite{BaBar-CP-Bpipi}),}\\
+1.00 \pm 0.21 \pm 0.07 & \mbox{(Belle \cite{Belle-CP-Bpipi}).}
\end{array}
\right.
\end{equation}
While these data differ from the ones used in \cite{BFRS-PRL,BFRS-BIG}, 
the averages that are relevant for us changed only marginally as seen in 
Table~\ref{tab:Bpipi-input}. Since their physical interpretation is in 
impressive accordance with the picture of the SM, as we will see below, 
we expect that the experimental results will stabilize around these numbers.

\boldmath
\subsection{Extraction of the Hadronic Parameters}
\unboldmath
If we assume that $\gamma$ and $\phi_d$ are known, (\ref{Rpipi-gen}) 
and (\ref{CP-Bpipi-gen}) allow us to convert the experimental
results for $R_{+-}^{\pi\pi}$,  $R_{00}^{\pi\pi}$ and
${\cal A}_{\rm CP}^{\rm dir}(B_d\to\pi^+\pi^-)$, 
${\cal A}_{\rm CP}^{\rm mix}(B_d\to\pi^+\pi^-)$ into values 
of $(d,\theta)$ and $(x,\Delta)$. Using the most recent results for 
the mixing-induced CP violation of the ``golden'' decay 
$B_d\to J/\psi K_{\rm S}$ (and related channels) obtained by the 
BaBar \cite{BaBar-s2b} and Belle collaborations \cite{Belle-s2b}, 
which correspond to the following new world average \cite{HFAG}:
\begin{equation}
\sin\phi_d=0.725 \pm 0.037, 
\end{equation}
we obtain
\begin{equation}\label{phi-d-range}
\phi_d=\left(46.5^{+3.2}_{-3.0}\right)^\circ,
\end{equation}
in excellent agreement with the picture of the SM \cite{CKM-Book}, and
with $\phi_d=\left(47\pm4\right)^\circ$ used in \cite{BFRS-PRL,BFRS-BIG}. 
It should be noted that we have neglected a second allowed solution for 
$\phi_d$ around $133^\circ$ in (\ref{phi-d-range}), which was analysed
in detail in \cite{Fl-Ma,FIM}. This possibility is now disfavoured by 
the data for CP violation in $B_d\to D^{(\ast)\pm}\pi^\mp$ decays 
\cite{RF-BDpi}, our previous $B\to\pi\pi,\pi K$ analysis \cite{BFRS-BIG}, 
and the first direct experimental result of the BaBar collaboration for 
the sign of $\cos\phi_d$. The latter follows from the measurement of the 
CP-violating observables of the time-dependent 
$B^0_d\to J/\psi[\to\ell^+\ell^-] K^{\ast}[\to \pi^0K_{\rm S}]$ 
angular distribution \cite{BaBar-c2b,CKM-fitters} 
(performing a similar analysis of the Belle data, it was, however, 
not possible to put constraints on the sign of 
$\cos\phi_d$ \cite{Belle-c2b}). Concerning $\gamma$, we assume the range
\begin{equation}\label{gamma-range}
\gamma=(65\pm7)^\circ,
\end{equation}
in accordance with the SM picture.

\begin{table}
\vspace{0.4cm}
\begin{center}
\begin{tabular}{|c||c|c|c|}
\hline
Parameter & EWPs included & \quad EWPs neglected & Previous analysis 
\\ \hline
 \hline
$d$  & $0.51^{+0.26}_{-0.20}$  &  $0.51^{+0.26}_{-0.20}$   & 
$0.48^{+0.35}_{-0.22}$ \\
\hline
$\theta$   &  $\left(140^{+14}_{-18}\right)^\circ$  &
$\left(140^{+14}_{-18}\right)^\circ$
   &  $+\left(138^{+19}_{-23}\right)^\circ$
\\ \hline
 \hline
$x$ &  $1.15^{+0.18}_{-0.16}$   &  $1.13^{+0.17}_{-0.16}$  & 
$1.22^{+0.26}_{-0.21}$ \\
\hline
$\Delta$  & $-\left(59^{+19}_{-26}\right)^\circ$   &
$-\left(57^{+20}_{-30}\right)^\circ$
   &  $-\left(71^{+19}_{-26}\right)^\circ$
\\ \hline
\end{tabular}
\end{center}
\caption[]{The hadronic parameters characterizing the $B\to\pi\pi$ 
system, extracted from the data summarized in Table~\ref{tab:Bpipi-input}
as explained in the text.}\label{tab:Bpipi-hadr}
\end{table}

If we complement now the experimental results summarized in
Table~\ref{tab:Bpipi-input} with (\ref{phi-d-range}) and 
(\ref{gamma-range}), we obtain the values of the hadronic parameters
collected in Table~\ref{tab:Bpipi-hadr}. In the numbers of the 
``EWPs included'' column, the EW penguins are taken into account
through the SM expressions in (\ref{B+pi+pi0EWP})--(\ref{Bdpi0pi0EWP}), 
in contrast to the ``EWPs neglected'' column. For the purpose of comparison, 
we give also the results of our previous analysis \cite{BFRS-PRL,BFRS-BIG}, 
where the EW penguin diagrams to the $B\to\pi\pi$ decays were neglected 
as well. We observe that the values of the hadronic parameters changed
only marginally through the new data, and that the impact of the 
EW penguin topologies on the extraction of $(x,\Delta)$ is in fact
small, as we anticipated. Note that the determination of $(d,\theta)$ 
is independent of the EW penguin effects.

\begin{figure}
\vspace*{0.3truecm}
\begin{center}
\includegraphics[width=10cm]{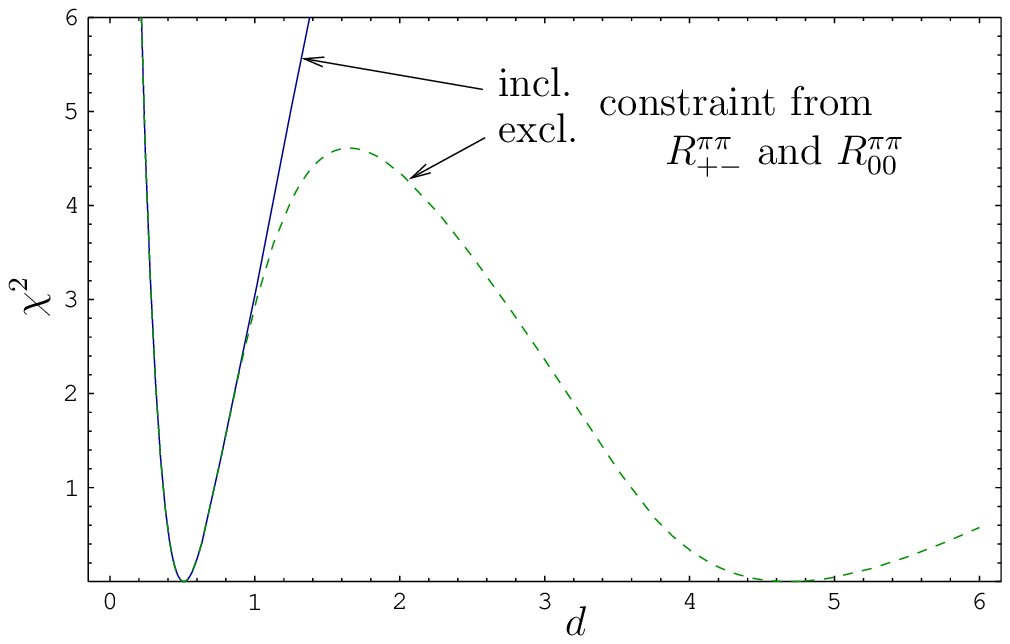}
\end{center}
\caption{$\chi^2$ of a fit to 
${\cal A}_{\rm CP}^{\rm dir}(B_d\to\pi^+\pi^-)$ and 
${\cal A}_{\rm CP}^{\rm mix}(B_d\to\pi^+\pi^-)$ with 
(solid) and without (dashed) a simultaneous fit to $R_{+-}^{\pi\pi}$ 
and $R_{00}^{\pi\pi}$.}\label{fig:d-ambig-res}
\end{figure}

As we discussed in terms of contours in the $\theta$--$d$ plane in 
\cite{BFRS-BIG}, the extraction of $(d,\theta)$ from 
${\cal A}_{\rm CP}^{\rm dir}(B_d\to\pi^+\pi^-)$ and 
${\cal A}_{\rm CP}^{\rm mix}(B_d\to\pi^+\pi^-)$ is affected by 
a twofold discrete ambiguity. However, imposing, in addition, the 
information provided by $R_{+-}^{\pi\pi}$ and $R_{00}^{\pi\pi}$,
we are only left with a single solution. A similar observation was 
subsequently also made by the authors of \cite{ALP-Bpipi}. In 
Fig.~\ref{fig:d-ambig-res}, we show the corresponding $\chi^2$ plot
for the determination of $d$ in order to illustrate this feature. 
For the resulting value of $(d,\theta)$, we obtain again a twofold 
solution for $(x,\Delta)$. However, this ambiguity can be resolved 
through the analysis of the $B\to\pi K$ system \cite{BFRS-BIG}, 
yielding the solution listed in Table~\ref{tab:Bpipi-hadr}.

The extraction of the hadronic parameters $(d,\theta)$ and $(x,\Delta)$
discussed above relies only on the isospin flavour symmetry of strong 
interactions, takes isospin-breaking effects through EW penguin processes 
into account, and is essentially theoretically clean. Similar analyses
were recently performed by several authors (see, for instance, 
\cite{BPRS,ALP-Bpipi,CGRS,He:2004ck}), who confirmed the picture found in
\cite{BFRS-PRL,BFRS-BIG}; the main differences between the various 
numerical results are due to the use of different input data. 

The hadronic parameters in Table~\ref{tab:Bpipi-hadr} allow us also 
to determine
\begin{equation}
\left[\frac{P}{T+C}\right]e^{i\gamma}=
-\left[\frac{de^{i\theta}}{1+xe^{i\Delta}}\right]=\frac{1}{R_b}
\left[\frac{{\cal P}_{tc}}{{\cal T}+{\cal C}}\right],
\end{equation}
yielding
\begin{equation}\label{P-T-rat1}
\left[\frac{P}{T+C}\right]e^{i\gamma}=
\left(0.27^{+0.10}_{-0.08}\right) \times e^{-i(8^{+18}_{-13})^\circ}.
\end{equation}
The experimental range for $R_b$ in (\ref{Rb-def}) implies then
\begin{equation}\label{Ptc-det}
\frac{{\cal P}_{tc}}{{\cal T}+{\cal C}}=
\left(0.10^{+0.04}_{-0.03}\right) \times e^{-i(8^{+18}_{-13})^\circ}.
\end{equation}
These values refer to the case, where the EW penguins are included
as in the SM; in the case of NP in the EW penguin sector, (\ref{P-T-rat1})
and (\ref{Ptc-det}) change in a negligible manner. For further 
discussions of these parameters, we refer the reader to 
\cite{BFRS-BIG,FR-I}.

\boldmath
\subsection{Theoretical Picture}
\unboldmath
It is instructive to compare Table~\ref{tab:Bpipi-hadr} with theoretical 
predictions. Concerning the ``QCD factorization approach'' (QCDF) 
\cite{BBNS1}, the most recent analysis was performed in \cite{BuSa}, where 
hadronic parameters $(r,\phi)$ were 
introduced,\footnote{These quantities should {\it not} be confused
with our $B\to\pi K$ parameters introduced in Section~\ref{sec:BpiK}.} 
which are related to $(d,\theta)$ through
\begin{equation}
d=\frac{r}{R_b}, \quad \theta=\phi-\pi.
\end{equation}
Using now the reference prediction for $r$ and $\phi$ in
QCDF given in \cite{BuSa}, $r=0.107\pm0.031$ and 
$\phi=\left(8.6\pm14.3\right)^\circ$, as well as the value of $R_b$ 
in (\ref{Rb-def}), we obtain
\begin{equation}\label{QCDF-pred}
\left.d\right|_{\rm QCDF}=0.29\pm0.09, \quad 
\left.\theta\right|_{\rm QCDF}=-\left(171.4\pm14.3\right)^\circ.
\end{equation}
On the other hand, the application of the ``perturbative hard-scattering
approach'' (PQCD) \cite{PQCD} yields the following prediction \cite{KeSa}:
\begin{equation}\label{PQCD-pred}
\left.d\right|_{\rm PQCD}=0.23^{+0.07}_{-0.05}, \quad 
+139^\circ < \left.\theta\right|_{\rm PQCD} < +148^\circ.
\end{equation}
We observe that the results for $d$ are in agreement with each other, 
but significantly smaller than the values given in 
Table~\ref{tab:Bpipi-hadr}. On the other hand, the PQCD picture for 
the strong phase $\theta$ is in accordance with the data, whereas QCDF 
favours a smaller phase with the {\it opposite} sign. For recent 
analyses using the framework of the ``soft collinear effective 
theory'' (SCET) \cite{SCET}, we refer the reader to \cite{BPRS,FeHu}.

Consequently, the theoretical attempts to calculate $d$ and $\theta$ from
first principles are not in accordance with the values following from the
SM interpretation of the current experimental data. This feature is already 
a challenge for several years (for earlier discussions, see, for instance,
\cite{Fl-Ma,RF-Bpipi}), and is now complemented by the measurement of
the $B_d\to\pi^0\pi^0$ channel with a rate that is significantly larger
than the one favoured in QCD factorization \cite{Be-Ne}. Unless the 
data will change in a dramatic manner, we have therefore to deal with
large non-factorizable effects. This conclusion is in agreement with
our previous one \cite{BFRS-PRL,BFRS-BIG}, and the conclusions drawn 
in \cite{BPRS,ALP-Bpipi,CGRS}.

\boldmath
\subsection{Prediction of the CP-Violating $B_d\to\pi^0\pi^0$
Observables}
\unboldmath
Having the hadronic parameters $(d,\theta)$ and $(x,\Delta)$ at hand, we 
may predict the CP-violating observables of the decay $B_d\to\pi^0\pi^0$,
which take the following generic form:
\begin{equation}\label{CP-Bpi0pi0-gen}
{\cal A}_{\rm CP}^{\rm dir}(B_d\to \pi^0\pi^0)=
H_1(d,\theta,x,\Delta;\gamma), \quad
{\cal A}_{\rm CP}^{\rm mix}(B_d\to \pi^0\pi^0)=
H_2(d,\theta,x,\Delta;\gamma,\phi_d);
\end{equation}
explicit expressions can be found in \cite{BFRS-BIG}. The conceptual
improvement with respect to our previous analysis is that we take
again the EW penguin contributions into account. Complementing the 
values in Table~\ref{tab:Bpipi-hadr} with (\ref{phi-d-range}) and 
(\ref{gamma-range}), we obtain the SM predictions
\begin{eqnarray}\label{ACP-dir-Bdpi0pi0}
\left.{\cal A}_{\rm CP}^{\rm dir}(B_d\to \pi^0\pi^0)\right|_{\rm SM}&=&
-0.28^{+0.37}_{-0.21}\,,\\
\label{ACP-mix-Bdpi0pi0}
\left.{\cal A}_{\rm CP}^{\rm mix}(B_d\to \pi^0\pi^0)\right|_{\rm SM}&=&
-0.63^{+0.45}_{-0.41}\,,
\end{eqnarray}
which are in good agreement with our previous numbers,
${\cal A}_{\rm CP}^{\rm dir}(B_d\to\pi^0\pi^0)=-0.41^{+0.35}_{-0.17}$ and 
${\cal A}_{\rm CP}^{\rm mix}(B_d\to\pi^0\pi^0)=-0.55^{+0.43}_{-0.45}$. 
We may now confront (\ref{ACP-dir-Bdpi0pi0}) with the first experimental 
results for the direct CP violation in $B_d\to\pi^0\pi^0$ that were 
recently reported by the BaBar and Belle collaborations:
\begin{equation}\label{ACP-dir-Bdpi0pi0-exp}
{\cal A}_{\rm CP}^{\rm dir}(B_d\to \pi^0\pi^0)=
\left\{
\begin{array}{ll}
-\left(0.12\pm0.56\pm0.06\right) & \mbox{(BaBar \cite{BaBar-Bpi0pi0}),}\\
-\left(0.43\pm0.51\,^{+0.17}_{-0.16}\right) & 
\mbox{(Belle \cite{Belle-Bpi0pi0-new}),}
\end{array}
\right.
\end{equation}
yielding the average of
\begin{equation}\label{ACP-dir-Bdpi0pi0-av}
{\cal A}_{\rm CP}^{\rm dir}(B_d\to \pi^0\pi^0)=-(0.28\pm0.39).
\end{equation}
Although the current errors are still large, the agreement between
(\ref{ACP-dir-Bdpi0pi0}) and (\ref{ACP-dir-Bdpi0pi0-av}) is 
very encouraging. We look forward to having more accurate data 
available. As we noted and illustrated in \cite{BFRS-BIG}, the measurement
of one of the CP-violating $B_d\to\pi^0\pi^0$ observables allows the
determination of $\gamma$.

\boldmath
\section{The $B\to\pi K$ System}\label{sec:BpiK}
\unboldmath
\setcounter{equation}{0}
\subsection{Preliminaries}\label{ssec:BpiK-pre}
The $B\to\pi K$ system consists of the four decay modes
$B^0_d\to\pi^-K^+$, $B^+\to\pi^+K^0$, $B^+\to\pi^0K^+$ and
$B^0_d\to\pi^0K^0$, which are governed by QCD penguin processes
\cite{RF-Phys-Rep}. A key difference between these transitions 
is due to EW penguin topologies: in the case of the 
former two channels, these may only contribute in colour-suppressed
form and are hence expected to play a minor r\^ole, whereas EW
penguins have a significant impact on the latter two transitions
thanks to colour-allowed contributions.\footnote{The neutral pions
can be emitted directly in these colour-allowed EW
penguin topologies.} In Table~\ref{tab:BpiK-input1},
we have summarized the current experimental status of the 
CP-averaged $B\to\pi K$ branching ratios. Following \cite{BFRS-PRL,BFRS-BIG},
it is possible to fix the hadronic $B\to\pi K$ parameters through their 
$B\to\pi\pi$ counterparts $(d,\theta)$ and $(x,\Delta)$. To this end,
we have to use the following working hypothesis:
\begin{itemize}
\item[i)] $SU(3)$ flavour symmetry of strong interactions.
\item[ii)] Neglect of penguin annihilation and exchange topologies. 
\end{itemize}
Concerning i), we include the factorizable $SU(3)$-breaking corrections
and perform internal consistency checks to probe non-factorizable 
$SU(3)$-breaking effects; the current data do not indicate large
corrections of this kind. Assumption ii) can be tested with the
help of $B_d\to K^+K^-$ and $B_s\to \pi^+\pi^-$ decays, where the
current experimental $B$-factory bounds for the former channel do 
not indicate any anomalous behaviour. In particular at LHCb, where 
also $B_s\to \pi^+\pi^-$ will be accessible, it should be possible
to explore the penguin annihilation and exchange topologies in a
much more stringent manner.

\begin{table}
\vspace{0.4cm}
\begin{center}
\begin{tabular}{|c||c|c|}
\hline
Quantity & This work & Previous analysis 
\\ \hline
 \hline
$\mbox{BR}(B_d\to\pi^\mp K^\pm)/10^{-6}$ & $18.2\pm0.8$ & $18.2\pm0.8$
\\ \hline
$\mbox{BR}(B^\pm\to\pi^\pm K)/10^{-6}$ & $24.1\pm1.3$ & $21.8 \pm 1.4$
\\ \hline
$\mbox{BR}(B^\pm\to\pi^0K^\pm)/10^{-6}$ & $12.1\pm0.8$ & $12.5\pm 1.1$
\\ \hline
$\mbox{BR}(B_d\to\pi^0K)/10^{-6}$ & $11.5\pm1.0$ & $11.7 \pm 1.4$
\\ \hline
 \hline
$R$ & $0.82\pm0.06$ & $0.91\pm0.07$
\\ \hline
$R_{\rm c}$ & $1.00\pm0.08$ & $1.17\pm0.12$
\\ \hline
$R_{\rm n}$ & $0.79\pm0.08$ & $0.76\pm0.10$
\\ \hline
\end{tabular}
\end{center}
\caption[]{The current status of the CP-averaged $B\to\pi K$ 
branching ratios, with averages taken from \cite{HFAG}, and 
comparison with the picture of our previous analysis 
\cite{BFRS-PRL,BFRS-BIG}. For completeness, we give also the
values of the ratios $R$, $R_{\rm c}$ and $R_{\rm n}$ introduced
in (\ref{R-def}), (\ref{Rc-def}) and (\ref{Rn-def}), where  
$R$ refers again to $\tau_{B^+}/\tau_{B^0_d}=1.086\pm0.017$ 
\cite{PDG}.}\label{tab:BpiK-input1}
\end{table}

\boldmath
\subsection{Direct CP Violation in 
$B_d\to\pi^\mp K^\pm$}\label{ssec:dir-CPV}
\unboldmath
\subsubsection{Experimental Picture}
The most important new experimental development in the $B\to\pi K$
sector is the observation of direct CP violation in $B_d\to\pi^\mp K^\pm$
decays, which could eventually be established this summer by the
BaBar \cite{BaBar-CP-dir-obs} and Belle \cite{Belle-CP-dir-obs} 
collaborations. These measurements complement the observation of
direct CP violation in the neutral kaon system by the NA48 (CERN) 
\cite{NA48-obs} and KTeV (FNAL) \cite{KTEV-obs} collaborations,
where this phenomenon is described by the famous observable 
$\mbox{Re}(\varepsilon'/\varepsilon)$; the world average taking 
the final results of these experiments \cite{NA48-final,KTEV-final} 
into account is given by 
$\mbox{Re}(\varepsilon'/\varepsilon)=(16.6\pm1.6)\times10^{-4}$. 
For recent theoretical overviews of $\mbox{Re}(\varepsilon'/\varepsilon)$, 
see \cite{BuJa,Pich:2004ee}.

In the case of $B_d\to\pi^\mp K^\pm$, direct CP violation is 
characterized by the asymmetry
\begin{equation}\label{ACPdir-exp}
{\cal A}_{\rm CP}^{\rm dir}(B_d\to\pi^\mp K^\pm)\equiv
\frac{\mbox{BR}(B^0_d\to\pi^-K^+)-
\mbox{BR}(\bar B^0_d\to\pi^+K^-)}{\mbox{BR}(B^0_d\to\pi^-K^+)+
\mbox{BR}(\bar B^0_d\to\pi^+K^-)},
\end{equation}
which is now measured by the BaBar and Belle collaborations with
the following results:\footnote{Note the different sign conventions!}
\begin{equation}\label{ACPdir-exps}
{\cal A}_{\rm CP}^{\rm dir}(B_d\to\pi^\mp K^\pm)=
\left\{
\begin{array}{ll}
+0.133 \pm 0.030 \pm 0.009 & \mbox{(BaBar \cite{BaBar-CP-dir-obs}),}\\
+0.101 \pm 0.025 \pm 0.005  & \mbox{(Belle \cite{Belle-CP-dir-obs}).}
\end{array}
\right.
\end{equation}
We observe that these numbers are nicely consistent with each other.
They correspond to the following average \cite{HFAG}:
\begin{equation}\label{ACPdir-av}
{\cal A}_{\rm CP}^{\rm dir}(B_d\to\pi^\mp K^\pm)=+0.113\pm 0.019,
\end{equation}
establishing the direct CP violation in $B_d\to\pi^\mp K^\pm$ decays
at the $5.9 \sigma$ level.

\subsubsection{Confrontation with Theory}\label{ACP-dir-theo1}
Let us now follow the strategy developed in \cite{BFRS-PRL,BFRS-BIG}
to confront the direct CP asymmetry of the $B_d^0\to\pi^- K^+$ channel
with theoretical considerations. The corresponding decay amplitude
can be written as 
\begin{equation}\label{Bdpi-K+}
A(B^0_d\to\pi^-K^+)=P'\left[1-re^{i\delta}e^{i\gamma}\right],
\end{equation}
with
\begin{equation}\label{P-prime}
P'\equiv\left(1-\frac{\lambda^2}{2}\right)A\lambda^2({\cal P}_t'-{\cal P}_c')
\end{equation}
and
\begin{equation}\label{r-def}
re^{i\delta}\equiv\left(\frac{\lambda^2R_b}{1-\lambda^2}
\right)\left[\frac{{\cal T}'-({\cal P}_t'-{\cal P}_u')}{{\cal P}_t'-
{\cal P}_c'}\right],
\end{equation}
yielding
\begin{equation}\label{ACPdir-r}
{\cal A}_{\rm CP}^{\rm dir}(B_d\to\pi^\mp K^\pm)=
\frac{2r\sin\delta\sin\gamma}{1-2r\cos\delta\cos\gamma+r^2}.
\end{equation}
The notation in (\ref{P-prime}) and (\ref{r-def})
is analogous to that used for the discussion of the
$B\to\pi\pi$ modes in Section~\ref{sec:Bpipi}; the primes remind 
us that we are dealing with $\bar b\to\bar s$ transitions. In
(\ref{ACPdir-r}), we can see nicely that 
${\cal A}_{\rm CP}^{\rm dir}(B_d\to\pi^\mp K^\pm)$ is induced through 
the interference between tree and QCD penguin topologies, with a 
CP-conserving strong phase difference $\delta$ and a CP-violating weak 
phase difference $\gamma$. 

If we use now the working hypothesis given in Subsection~\ref{ssec:BpiK-pre}, 
we obtain \cite{BFRS-PRL,BFRS-BIG}
\begin{equation}\label{r-det}
re^{i\delta}=\frac{\epsilon}{d}e^{i(\pi-\theta)}.
\end{equation}
This relation allows us to determine $(r,\delta)$ from the values of the 
hadronic $B\to\pi\pi$ parameters $(d,\theta)$ given in 
Table~\ref{tab:Bpipi-hadr}, with the following result:
\begin{equation}\label{r-del-det}
r=0.10^{+0.05}_{-0.04}, \quad \delta=+\left(39.6^{+17.7}_{-13.9}\right)^\circ.
\end{equation}
Having these parameters at hand, which refer to the range for $\gamma$
in (\ref{gamma-range}), we are in a position to calculate the direct 
CP asymmetry of $B_d\to\pi^\mp K^\pm$:
\begin{equation}\label{ACP-pred}
{\cal A}_{\rm CP}^{\rm dir}(B_d\to\pi^\mp K^\pm)=+0.127^{+0.102}_{-0.066},
\end{equation}
which should be compared with our previous prediction of 
$+0.140^{+0.139}_{-0.087}$ \cite{BFRS-BIG}. Looking at (\ref{ACPdir-av}), 
we observe that (\ref{ACP-pred}) is in nice agreement with the 
experimental result. In fact, in our previous analysis, which was
confronted with the experimental average of  
${\cal A}_{\rm CP}^{\rm dir}(B_d\to\pi^\mp K^\pm)=+0.095\pm0.028$, 
we advocated that this CP asymmetry should go up, in full accordance 
with the BaBar result in (\ref{ACPdir-exps}). Despite the large 
value of $\delta$, the rather small value of $r$ ensures that
this CP asymmetry does not take a value that is much larger than
the experimental ones. 

In the case of QCDF \cite{Be-Ne,BBNS-BpiK} and PQCD \cite{KeSa}, 
the following patterns arise:
\begin{equation}
\left.{\cal A}_{\rm CP}^{\rm dir}(B_d\to\pi^\mp K^\pm)\right|_{\rm QCDF}\sim
-(0.05\pm0.09),
\end{equation}
\begin{equation}
+0.13 \, \lsim \,
\left.{\cal A}_{\rm CP}^{\rm dir}(B_d\to\pi^\mp K^\pm)\right|_{\rm PQCD}
\, \lsim \, +0.22.
\end{equation}
Consequently, the QCDF picture is not in agreement with the experimental
result (\ref{ACPdir-av}), pointing in particular towards the 
{\it opposite} sign of the direct CP asymmetry. On the other hand, PQCD 
reproduces the sign correctly, but favours an asymmetry on the larger 
side. These features can also be seen with the help of (\ref{ACPdir-r}) 
and (\ref{r-det}) from the QCDF and PQCD predictions given in 
(\ref{QCDF-pred}) and (\ref{PQCD-pred}), respectively.

\subsubsection{Alternative Confrontation with Theory}\label{ssec:alt-conf}
Another direct confrontation of 
${\cal A}_{\rm CP}^{\rm dir}(B_d\to\pi^\mp K^\pm)$
with theory is provided by the following SM relation, 
which can be derived with the help of assumptions i) and ii) specified 
in Subsection~\ref{ssec:BpiK-pre} \cite{RF-Bpipi,Deshpande:1994ii,RF-BsKK}:
\begin{equation}\label{CP-rel}
\epsilon H \equiv 
\underbrace{\left(\frac{f_K}{f_\pi}\right)^2\left[\frac{\mbox{BR}
(B_d\to\pi^+\pi^-)}{\mbox{BR}(B_d\to\pi^\mp K^\pm)}
\right]}_{\mbox{0.38$\pm$0.04}}=
\underbrace{-\left[\frac{{\cal A}_{\rm CP}^{\rm dir}(B_d\to\pi^\mp 
K^\pm)}{{\cal A}_{\rm CP}^{\rm dir}(B_d\to\pi^+\pi^-)}
\right]}_{\mbox{0.31$\pm$0.11}}.
\end{equation}
Here, we have introduced the parameter
\begin{equation}
\epsilon\equiv\frac{\lambda^2}{1-\lambda^2}=0.053,
\end{equation}
and the ratio $f_K/f_\pi=160/131$ of the kaon and pion decay constants 
takes the factorizable $SU(3)$-breaking corrections into account. 
In (\ref{CP-rel}), we have also indicated the current experimental 
results, and observe that this relation is nicely
satisfied within the current experimental uncertainties. This feature
give us further confidence in our working assumptions, in addition to
the agreement between (\ref{ACPdir-av}) and (\ref{ACP-pred}).

\begin{figure}
\vspace*{0.3truecm}
\begin{center}
\includegraphics[width=13cm]{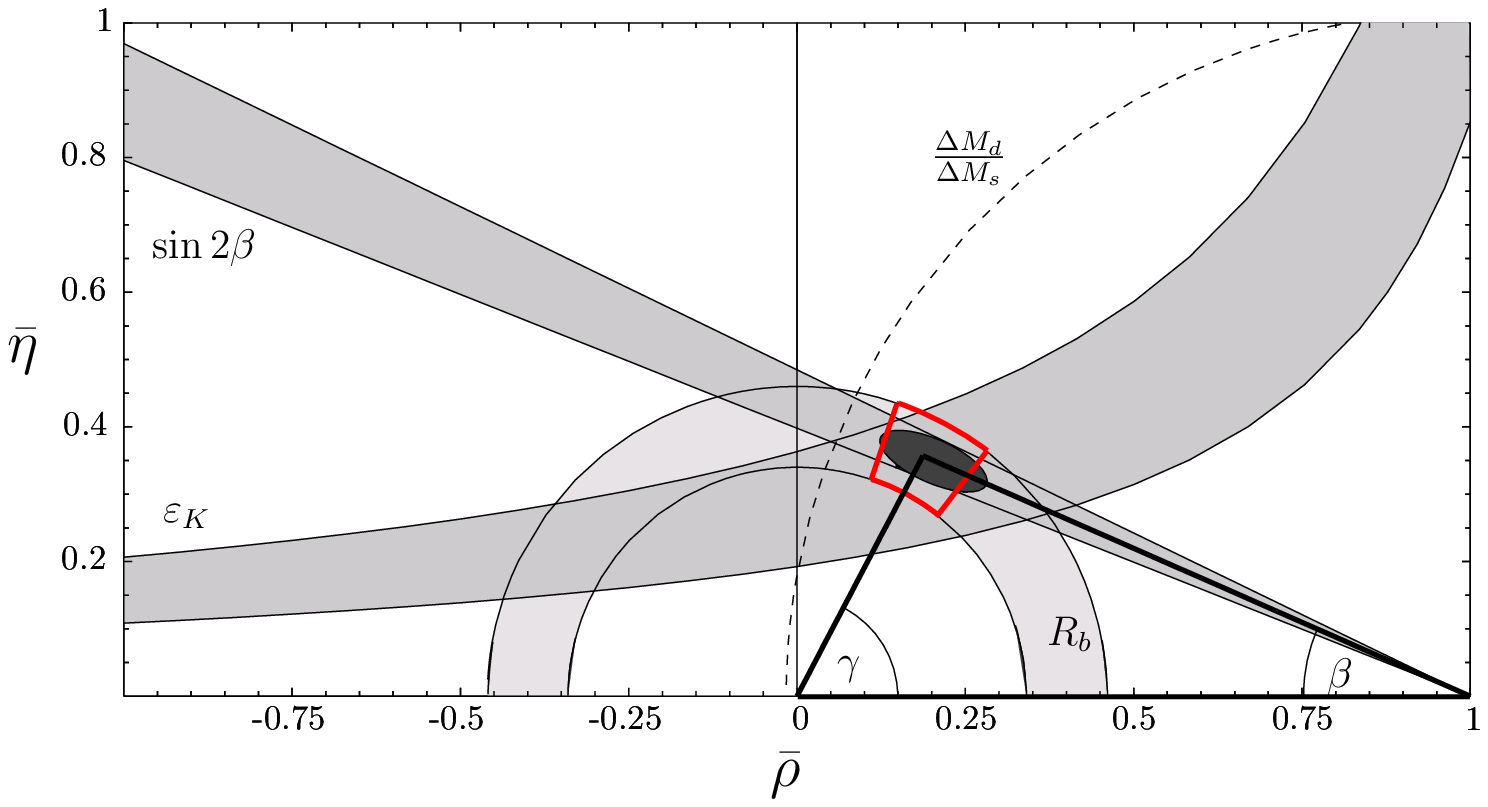}
\end{center}
\vspace*{-0.5truecm}
\caption{Illustration of the value of $\gamma$ following from the 
CP-violating $B_d\to\pi^+\pi^-$ observables and the data for the
$B_d\to\pi^\mp K^\pm$ decays in the $\bar\rho$--$\bar \eta$ plane 
and comparison with the other constraints for the UT, as discussed
in \cite{BSU}. The shaded dark ellipse is the result of the UT fit,
while the quadrangle corresponds to the second value of
$\left.\gamma\right|_{\rm BR}$ in (\ref{gamma}) and the $R_b$ 
constraint.}\label{fig:UT}
\end{figure}

\subsubsection{Implications for the UT}\label{ssec:UT-det}
The quantity $H$ introduced in (\ref{CP-rel}) can be written as
follows:
\begin{equation}\label{H-fct}
H=G_3(d,\theta;\gamma).
\end{equation}
If we now complement $H$ with the CP-violating observables
${\cal A}_{\rm CP}^{\rm dir}(B_d\to\pi^+\pi^-)$ and 
${\cal A}_{\rm CP}^{\rm mix}(B_d\to\pi^+\pi^-)$, which take the
general form in (\ref{CP-Bpipi-gen}), and use the experimental
result for $\phi_d$ in (\ref{phi-d-range}), we are in a position 
to determine $\gamma$ and $(d,\theta)$
\cite{Fl-Ma,FIM,RF-Bpipi,RF-BsKK}. In addition to the expression 
involving the CP-averaged $B_d\to\pi^+\pi^-$ and $B_d\to\pi^\mp K^\pm$ 
branching ratios in (\ref{CP-rel}), the corresponding direct 
CP asymmetries provide an alternative avenue for the determination
of $H$, which is theoretically more favourable as far as the 
$SU(3)$-breaking corrections are concerned, but is currently 
affected by larger experimental uncertainties. The corresponding values 
of $H$ are given as follows:
\begin{equation}
\left.H\right|_{\rm BR}=7.2\pm 0.7, \quad
\left.H\right|_{{\cal A}_{\rm CP}^{\rm dir}}=5.9\pm 2.1.
\end{equation}
Complementing them with the CP-violating $B_d\to\pi^+\pi^-$ asymmetries in 
Table~\ref{tab:Bpipi-input}, we obtain the following solutions for $\gamma$:
\begin{equation}
\label{gamma}
\left.\gamma\right|_{\rm BR}=\left(39.6\,^{+5.8}_{-4.6} \right)^\circ \lor 
\left(63.3\,^{+7.7}_{-11.1}\right)^\circ, \quad
\left.\gamma\right|_{{\cal A}_{\rm CP}^{\rm dir}}=
\left(38.1\,^{+5.4}_{-5.6}\right)^\circ \lor 
\left(66.6\,^{+11.0}_{-11.1}\right)^\circ. 
\end{equation}
As we discussed in \cite{BFRS-BIG}, the twofold ambiguities arising in
these determinations can be lifted by using additional experimental 
information for $B^\pm\to \pi^\pm K$ and the other $B\to\pi\pi$ decays, 
thereby leaving us with the values around $65^\circ$. In Fig.~\ref{fig:UT}, 
we show the corresponding situation for the UT in the 
$\bar\rho$--$\bar \eta$ plane of the generalized Wolfenstein 
parameters \cite{BLO}, where we compare the values of $\gamma$ obtained 
above with the UT fit performed in \cite{BSU}. Using, in
addition, the range for the UT side $R_b$ in (\ref{Rb-def}), we may
also determine $\alpha$ and $\beta$, with the following results:
\begin{equation}
\left.\alpha\right|_{\rm BR}=\left(95.0\,^{+12.2}_{-8.2}\right)^\circ, \quad
\left.\alpha\right|_{{\cal A}_{\rm CP}^{\rm dir}}=
\left(91.7\,^{+12.0}_{-11.0}\right)^\circ,
\label{alpha-determ}\end{equation}
\begin{equation}
\left.\beta\right|_{\rm BR}=\left(21.6\,^{+2.6}_{-2.7}\right)^\circ, \quad
\left.\beta\right|_{{\cal A}_{\rm CP}^{\rm dir}}=
\left(21.7\,^{+2.5}_{-2.6}\right)^\circ.
\label{beta-determ}\end{equation}
The results for $\alpha$ are nicely consistent with those obtained from
the most recent data for $B\to\rho\rho,\rho\pi$ processes, as reviewed 
in \cite{ligeti}; a similar comment applies to the ranges for $\gamma$ 
following from decays of the kind $B^\pm\to D K^\pm$. Let us also 
emphasize that our results for $\beta$ are in excellent agreement with 
the SM relation $\phi_d=2\beta$. In this context, it is important to 
stress that actually $\phi_d$ -- and not $\beta$ itself -- enters our 
analysis as an input parameter.\footnote{The only place where actually 
$\beta$ enters is in (\ref{B+pi+pi0EWP}) and (\ref{Bdpi0pi0EWP}) to 
describe the tiny EW penguin effects in $B^+\to \pi^+\pi^0$ and 
$B^0_d\to\pi^0\pi^0$, respectively. However, $\beta$ does {\it not} 
enter the $B^0_d\to\pi^+\pi^-$ amplitude (\ref{Bdpi+pi-EWP}), which is 
the only relevant $B\to\pi\pi$ ingredient for our UT analysis.}
The determination of $\alpha$ and $\beta$ in (\ref{alpha-determ}) and
(\ref{beta-determ}) is therefore an important test of the consistency
of our approach (or of the SM relation $\phi_d=2\beta$). 

More refined determinations of $\gamma$ from the
CP-violating $B_d\to\pi^+\pi^-$ observables are provided by the
decay $B_s\to K^+K^-$ \cite{RF-BsKK}. This channel is already
accessible at run II of the Tevatron \cite{TEV-BOOK,CDF}, and 
can be fully exploited at LHCb \cite{LHC-BOOK,LHCb}. In 
Appendix~\ref{app:BsKK}, we collect the updated values of the 
SM predictions for the $B_s\to K^+K^-$ observables presented 
in \cite{BFRS-BIG}.

\boldmath
\subsection{The $B_d\to\pi^\mp K^\pm$, $B^\pm\to\pi^\pm K$
System}\label{ssec:BpiK-mix}
\unboldmath
\subsubsection{Experimental Picture}
The direct CP violation in $B_d\to\pi^\mp K^\pm$ decays provides 
valuable information and is perfectly consistent with the SM picture 
emerging from our strategy~\cite{BFRS-PRL,BFRS-BIG}. Let us now also
consider the CP-averaged $B_d\to\pi^\mp K^\pm$ rate. In order 
to analyse this quantity, it is useful to consider simultaneously
$B^\pm\to\pi^\pm K$ decays \cite{PAPIII,FM,GR-BpiK-97,defan},
and to introduce 
\begin{equation}\label{R-def}
R\equiv\left[\frac{\mbox{BR}(B_d^0\to\pi^- K^+)+
\mbox{BR}(\bar B_d^0\to\pi^+ K^-)}{\mbox{BR}(B^+\to\pi^+ K^0)+
\mbox{BR}(B^-\to\pi^- \bar K^0)}
\right]\frac{\tau_{B^+}}{\tau_{B^0_d}}.
\end{equation}
The common feature of the $B^0_d\to\pi^-K^+$ and $B^+\to\pi^+K^0$ decays
is that EW penguins may only contribute to them in colour-suppressed form, 
and are hence expected to play a minor r\^ole. As can be seen in 
Table~\ref{tab:BpiK-input1}, the experimental average for $R$ went down 
sizeably with respect to the situation of our previous analysis. This 
feature is essentially due to the most recent update of the CP-averaged 
$B^\pm\to\pi^\pm K$ branching ratio by the BaBar collaboration 
\cite{BaBar-BdKK-obs}, taking certain radiative corrections into 
account; similar effects are currently investigated by the Belle 
collaboration \cite{browder}. Consequently, the experimental picture is 
not yet settled (see also \cite{ligeti}). 

The last observable provided by the $B_d\to\pi^\mp K^\pm$, 
$B^\pm\to\pi^\pm K$ system is -- in addition to 
${\cal A}_{\rm CP}^{\rm dir}(B_d\to\pi^\mp K^\pm)$ and $R$ -- the
direct CP asymmetry of the $B^\pm\to\pi^\pm K$ modes:
\begin{equation}
{\cal A}_{\rm CP}^{\rm dir}(B^\pm\to\pi^\pm K)\equiv
\frac{\mbox{BR}(B^+\to\pi^+K^0)-
\mbox{BR}(B^-\to\pi^-\bar K^0)}{\mbox{BR}(B^+\to\pi^+K^0)+
\mbox{BR}(B^-\to\pi^-\bar K^0)}.
\end{equation}
The current experimental average is given as follows \cite{HFAG}:
\begin{equation}\label{ACP-B+pi+K-exp}
{\cal A}_{\rm CP}^{\rm dir}(B^\pm\to\pi^\pm K)=+0.020\pm0.034,
\end{equation}
and does not indicate any CP-violating effects in this channel.

\subsubsection{Confrontation with Theory}\label{ssec:R-puzzle}
In order to complement the $B^0_d\to\pi^-K^+$ amplitude in 
(\ref{Bdpi-K+}), we write
\begin{equation}\label{B+pi+K0}
A(B^+\to\pi^+K^0)=-P'\left[1+\rho_{\rm c}e^{i\theta_{\rm c}}e^{i\gamma}
\right],
\end{equation}
where $P'$ was defined in (\ref{P-prime}), and 
\begin{equation}\label{rho-c-def}
\rho_{\rm c}e^{i\theta_{\rm c}}\equiv\left(\frac{\lambda^2R_b}{1-\lambda^2}
\right)\left[\frac{{\cal P}_t'-
\tilde {\cal P}_u'-{\cal A}'}{{\cal P}_t'-{\cal P}_c'}\right].
\end{equation}
Here $\tilde {\cal P}_u'$ describes the penguins with internal up-quark 
exchanges contributing to the {\it charged} $B\to\pi K$ modes, and 
${\cal A}'$ is an annihilation topology. We arrive then straightforwardly
at the following expression for $R$:
\begin{equation}\label{R-expr}
R=\frac{1-2r\cos\delta\cos\gamma+r^2}{1+2\rho_{\rm c}
\cos\theta_{\rm c}\cos\gamma+\rho_{\rm c}^2},
\end{equation}
while the direct CP asymmetry of the $B^\pm\to\pi^\pm K$ modes 
is given by
\begin{equation}\label{ACP-B+pi+K-expr}
{\cal A}_{\rm CP}^{\rm dir}(B^\pm\to\pi^\pm K)=
-\left[\frac{2\rho_{\rm c}\sin\theta_{\rm c}\sin\gamma}{1+2\rho_{\rm c}
\cos\theta_{\rm c}\cos\gamma+\rho_{\rm c}^2}\right].
\end{equation}

Let us first assume that $\rho_{\rm c}$ can be neglected, as is usually
done for the analysis of the $B_d\to\pi^\mp K^\pm$, $B^\pm\to\pi^\pm K$
system. This approximation corresponds to a vanishing value of
(\ref{ACP-B+pi+K-expr}), which is in accordance with (\ref{ACP-B+pi+K-exp}).
In view of the rather small experimental value of $R$, it is interesting
(see also \cite{ligeti}) to return to the bounds on $\gamma$ that can be 
obtained with the help of 
\begin{equation}\label{Rmin}
\sin^2\gamma\leq R,
\end{equation}
provided $R$ is measured to be smaller than 1 \cite{FM}. Using the value 
of $R=0.82\pm0.06$ in Table~\ref{tab:BpiK-input1}, we obtain the upper 
bound 
\begin{equation}\label{FM-bound1}
\gamma\leq \left(64.9^{+4.8}_{-4.2}\right)^\circ,
\end{equation}
which is basically identical with the range for $\gamma$ in 
(\ref{gamma-range}). 

In analogy to the prediction in (\ref{ACP-pred}), the hadronic parameters 
in (\ref{r-del-det}) allow us also to calculate $R$, with the following 
result: 
\begin{equation}\label{R-pred1}
R=0.943^{+0.028}_{-0.021},
\end{equation}
which is the update of $R=0.943^{+0.033}_{-0.026}$ given in 
\cite{BFRS-PRL,BFRS-BIG}. Comparing with the new 
experimental result in Table \ref{tab:BpiK-input1}, we observe that 
it favours a smaller value. 
Consequently, in the case of $R$, we encounter now a sizeable 
deviation of our prediction from the experimental average, whereas 
we obtain excellent agreement with the $B$-factory data for the 
CP-violating $B_d\to\pi^\mp K^\pm$ asymmetry. Moreover, also the
bound on $\gamma$ in (\ref{FM-bound1}) appears to be surprisingly
close to the SM range. Since $R$ may be affected by $\rho_{\rm c}$, 
whereas ${\cal A}_{\rm CP}^{\rm dir}(B_d\to\pi^\mp K^\pm)$ does 
{\it not} involve this hadronic parameter, it is therefore suggested
that $\rho_{\rm c}$ has actually a non-negligible impact on the
numerical analysis.

\boldmath
\subsubsection{A Closer Look at $\rho_{\rm c}$}\label{ssec:rhoc-analysis}
\begin{figure}
\begin{center}
\includegraphics[width=11.5cm]{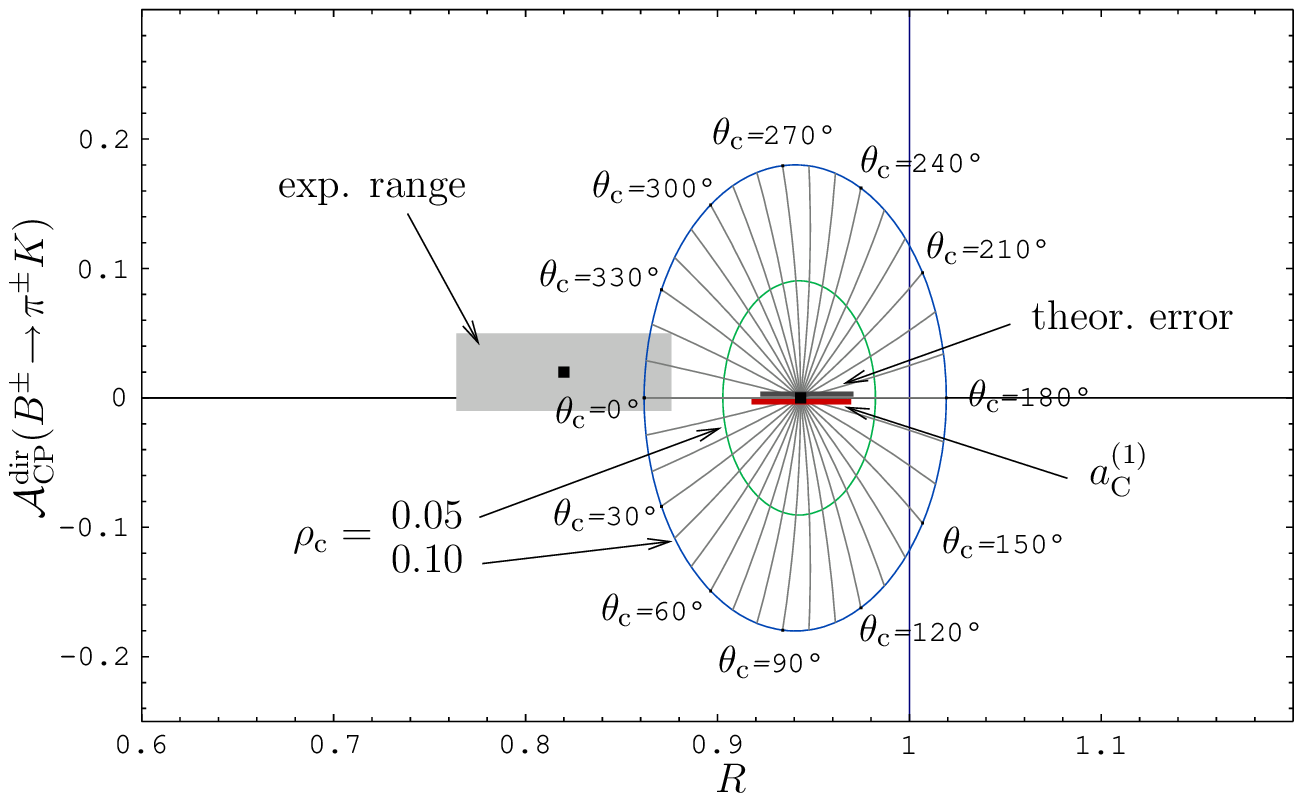}
\end{center}
\vspace*{-0.3truecm}
\caption{The situation in the 
$R$--${\cal A}_{\rm CP}^{\rm dir}(B^\pm\!\to\!\pi^\pm K)$ plane. 
We show contours for $\rho_{\rm c}=0.05$ and $\rho_{\rm c}=0.10$, with 
$\theta_{\rm c} \in [0^\circ,360^\circ]$. The experimental ranges for $R$ 
and ${\cal A}_{\rm CP}^{\rm dir}(B^\pm\!\to\!\pi^\pm K)$ and our 
theoretical prediction are indicated in grey. The second error bar
beneath the theoretical prediction (almost identical in size to
the first one) indicates the variation of $R$ if colour-suppressed
EW penguins are taken into account.\label{fig:R-Adir}}
\end{figure}

\unboldmath
In addition to $R$, the parameter $\rho_{\rm c}$ enters also the direct 
CP asymmetry of the $B^\pm\to \pi^\pm K$ decays. It is interesting to 
illustrate these effects in the 
$R$--${\cal A}_{\rm CP}^{\rm dir}(B^\pm\!\to\!\pi^\pm K)$ plane.
To this end, we use (\ref{R-expr}) with the central values of the 
hadronic parameters in (\ref{r-del-det}), and (\ref{ACP-B+pi+K-expr})
to calculate the contours for $\rho_{\rm c}=0.05$ and $0.10$ shown 
in Fig.~\ref{fig:R-Adir}. We observe that for $\rho_{\rm c}=0.05$ and 
$-30^\circ \leq \theta_{\rm c} \lsim 0^\circ$ the $1 \sigma$ ranges
of experiment and theory (the contours only show the central value)
practically overlap, thereby resolving essentially the discrepancy 
between our theoretical prediction for $R$ and its most recent 
experimental value. Following Appendix~D.3 of \cite{BFRS-BIG}, we 
have also included a second error bar for our theoretical prediction 
that indicates the variation of $R$ if colour-suppressed EW penguins 
are taken into account at a rather prominent level of 
$a_{\rm C}^{(1)}=0.1$, with $\Delta_{\rm C}^{(1)}\in[0,360^\circ]$. 
We observe that, while the inclusion of these effects could also help 
to resolve the discrepancy, the impact of $\rho_{\rm c}$ is significantly 
more important.

After this encouraging observation, let us have a closer look at the status 
of $\rho_{\rm c}$. Access to this parameter is provided by the decay
$B^+\to K^+ \bar K^0$, which is related to $B^+\to \pi^+ K^0$ through
the interchange of all down and strange quarks, i.e.\ through the
$U$-spin flavour symmetry of strong interactions \cite{BFM,FKNP},
which is a subgroup of $SU(3)_{\rm F}$. Applying this symmetry, we 
may write (for a detailed discussion, see \cite{RF-Phys-Rep})
\begin{equation}\label{K-def}
K\equiv\left[\frac{1}{\epsilon R_{SU(3)}^2}\right]
\left[\frac{\mbox{BR}(B^\pm\to \pi^\pm K)}{\mbox{BR}(B^\pm\to K^\pm K)}
\right]=\frac{1+2\rho_{\rm c}\cos\theta_{\rm c}\cos\gamma+
\rho_{\rm c}^2}{\epsilon^2-2\epsilon\rho_{\rm c}\cos\theta_{\rm c}\cos\gamma+
\rho_{\rm c}^2},
\end{equation}
where $R_{SU(3)}$ describes $SU(3)$-breaking corrections. In factorization,
we obtain
\begin{equation}\label{R-SU3}
R_{SU(3)}=\left[\frac{M_B^2-M_\pi^2}{M_B^2-M_K^2}\right]
\left[\frac{F_{B\pi}(M_K^2;0^+)}{F_{BK}(M_K^2;0^+)}\right]=0.79,
\end{equation}
where the numerical value refers to the recent light-cone sum-rule
analysis performed in \cite{Ball}. The measurement of $K$ allows us to 
obtain the following allowed range for $\rho_{\rm c}$:
\begin{equation}\label{rhoc-bounds}
\frac{1-\epsilon\sqrt{K}}{1+\sqrt{K}}\leq\rho_{\rm c}\leq
\frac{1+\epsilon\sqrt{K}}{|1-\sqrt{K}|}.
\end{equation}
Using the most recent upper bound of 
$\mbox{BR}(B^\pm\to K^\pm K)<2.35 \times 10^{-6}$ (90\% C.L.) 
reported by the BaBar collaboration \cite{BaBar-BdKK-obs}, and
the measured value of $\mbox{BR}(B^\pm\to \pi^\pm K)$ in
Table~\ref{tab:BpiK-input1}, this relation implies
\begin{equation}\label{rhoc-range1}
\rho_{\rm c}<0.13.
\end{equation}

\begin{figure}
\begin{center}
\includegraphics[width=11.5cm]{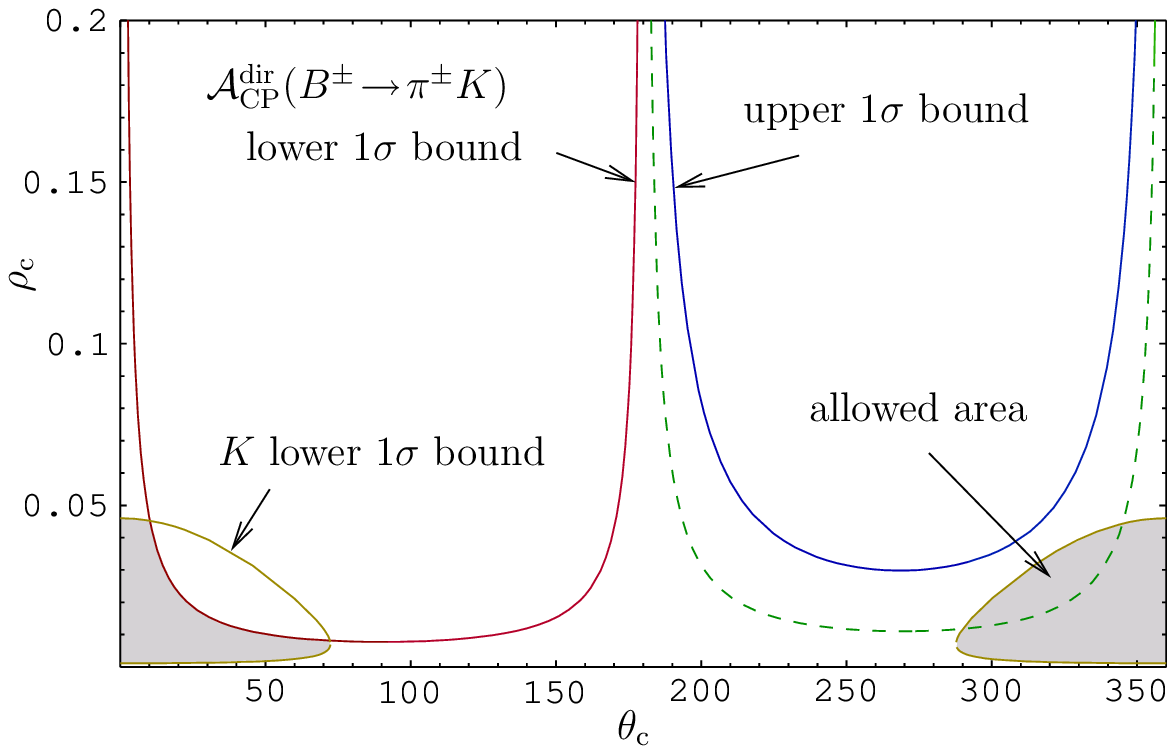}
\end{center}
\vspace*{-0.3truecm}
\caption{The constraints in the $\theta_{\rm c}$--$\rho_{\rm c}$ plane
following from the current data for the CP-averaged $B^\pm\to K^\pm K$
branching ratio (parametrized by $K$) and the direct CP violation 
in $B^\pm\to\pi^\pm K$ decays as discussed in the text.}\label{fig:rhoc-thetac-dat}
\end{figure}

The neutral counterpart of $B^+\to K^+\bar K^0$, the $B^0_d\to K^0\bar K^0$ 
channel, was observed this summer by the BaBar collaboration 
\cite{BaBar-BdKK-obs}, with the CP-averaged branching ratio 
\begin{equation}
\mbox{BR}(B_d\to K^0\bar K^0)=\left(1.19^{+0.40}_{-0.35}\pm0.13\right)
\times 10^{-6},
\end{equation}
corresponding to a significance of $4.5\sigma$. This exciting measurement 
is the first {\it direct} experimental evidence for a $b\to d$ penguin 
process. Interestingly, it is in accordance with the lower SM bounds 
derived in \cite{FR-I}, which suggested that the discovery 
of this transition should actually be just ahead of us. Concerning
$B^+\to K^+\bar K^0$, there is an emerging signal at the $3.5\sigma$
level, which would correspond to
\begin{equation}\label{BR-B+K+K}
\mbox{BR}(B^\pm\to K^\pm K)=\left(1.45^{+0.53}_{-0.46}\pm0.11\right)
\times 10^{-6};
\end{equation}
in the SM, we expect a lower bound of $(1.69^{+0.21}_{-0.24})\times
10^{-6}$ \cite{FR-II}. Inserting the range in (\ref{BR-B+K+K}) 
into (\ref{rhoc-bounds}) yields
\begin{equation} \label{rhoc-bounds-num}
-0.008\pm0.008<\rho_{\rm c}<0.102\pm0.009.
\end{equation}

If we use the SM value of $\gamma$ in (\ref{gamma-range}), the 
measurement of $\mbox{BR}(B^\pm\to K^\pm K)$ provides even more 
information. In fact, (\ref{K-def}) allows us then to determine 
$\rho_{\rm c}$ as a function of $\theta_{\rm c}$ with the help of
\begin{equation}\label{rhoc-cont1}
\rho_{\rm c}=\tilde a\pm\sqrt{\tilde a^2-\tilde b},
\end{equation}
where
\begin{equation}
\tilde a\equiv\left[\frac{\epsilon K+1}{K-1}\right]
\cos\gamma\cos\theta_{\rm c},
\quad
\tilde b\equiv\frac{\epsilon^2 K-1}{K-1}.
\end{equation}
In analogy, the experimental result for ${\cal A}_{\rm CP}^{\rm dir}
(B^\pm\to\pi^\pm K)$ allows us to fix another contour in the
$\theta_{\rm c}$--$\rho_{\rm c}$ plane. Using (\ref{ACP-B+pi+K-expr}),
we obtain
\begin{equation}\label{rhoc-cont2}
\rho_{\rm c}=-\tilde c\pm\sqrt{\tilde c^2-1},
\end{equation}
with
\begin{equation}
\tilde c=
\left[1+ \frac{\tan\gamma\tan\theta_{\rm c}}{{\cal A}_{\rm CP}^{\rm dir}
(B^\pm\to\pi^\pm K)}\right]\cos\gamma\cos\theta_{\rm c}.
\end{equation}

In Fig.~\ref{fig:rhoc-thetac-dat}, we assume $\gamma=65^\circ$, and
confront these considerations with the experimental results in 
(\ref{ACP-B+pi+K-exp}) and (\ref{BR-B+K+K}), despite the fact that 
the latter branching ratio corresponds only to an emerging signal
for the $B^\pm\to K^\pm K$ channel. Since the corresponding lower
$1\sigma$ and central values of $\mbox{BR}(B^\pm\to K^\pm K)$ 
would be smaller than the lower bound derived in \cite{FR-II}, 
(\ref{rhoc-cont1}) would not have a physical solution for these
results. However, for values of $\mbox{BR}(B^\pm\to K^\pm K)$
larger than this bound, we obtain an expanding allowed region
in the $\theta_{\rm c}$--$\rho_{\rm c}$ plane, as shown in
Fig.~\ref{fig:rhoc-thetac-dat}. We also observe that 
${\cal A}_{\rm CP}^{\rm dir}(B^\pm\to\pi^\pm K)$ has a 
rather small impact on the overall allowed parameter space. It
is interesting to note that the data favour strong phases 
$\theta_{\rm c}$ around $0^\circ$ (and not around $180^\circ$), as 
is suggested by the general expression in (\ref{rho-c-def}). 

In the following, we will use $\rho_{\rm c}=0.05$ and 
$\theta_{\rm c}=0^\circ$, in agreement with (\ref{rhoc-bounds-num})
and the allowed region in Fig.~\ref{fig:rhoc-thetac-dat};
a more rigorous analysis will have to wait until the data for 
the $B^\pm\to K^\pm K$ decays will have improved. With 
these values and $\gamma=65^\circ$, we obtain
\begin{equation}\label{wc-range}
w_{\rm c}\equiv\sqrt{1+2\rho_{\rm c}\cos\theta_{\rm c}\cos\gamma+
\rho_{\rm c}^2}=1.022\,.
\end{equation}
As can be seen in (\ref{R-expr}), this quantity describes the impact 
of $(\rho_{\rm c},\theta_{\rm c})$ on $R$. In particular, the numerical 
value in (\ref{wc-range}) shifts the central value $R=0.943$ in 
(\ref{R-pred1}) accordingly to 0.903. We observe that $R$ moves actually
towards the experimental value through the impact of $\rho_{\rm c}$, 
thereby essentially resolving the discrepancy arising in
Subsection~\ref{ssec:R-puzzle}. 

To conclude the discussion of the $B_d\to\pi^\mp K^\pm$, 
$B^\pm\to\pi^\pm K$ system, let us emphasize that we can
accommodate the corresponding data in the SM by using additional
experimental information on $B^\pm\to K^\pm K$ decays, allowing
us to take the hadronic parameter $\rho_{\rm c}$ into account. 
The remaining small numerical difference in the analysis of $R$,
if confirmed by future data, could be due to (small) effects of 
colour-suppressed EW penguins, which enter $R$ as well 
\cite{BFRS-BIG,defan}, and/or the limitations of our working 
hypothesis specified in Subsection~\ref{ssec:BpiK-pre}. Moreover, 
we would also not be surprised to see the experimental value of 
$R$ moving up in the future.

\boldmath
\subsection{The Charged and Neutral $B\to\pi K$ Systems}\label{ssec:BpiK-CN}
\unboldmath
\subsubsection{Experimental Picture}\label{ssec:EXP-c-n}
Let us now turn to the decays $B^+\to\pi^0K^+$ and
$B^0_d\to\pi^0K^0$, where EW penguins enter in colour-allowed form.
In order to analyse these transitions, it is particularly useful to 
introduce the following ratios:
\begin{equation}\label{Rc-def}
R_{\rm c}\equiv2\left[\frac{\mbox{BR}(B^+\to\pi^0K^+)+
\mbox{BR}(B^-\to\pi^0K^-)}{\mbox{BR}(B^+\to\pi^+ K^0)+
\mbox{BR}(B^-\to\pi^- \bar K^0)}\right]
\end{equation}
\begin{equation}\label{Rn-def}
R_{\rm n}\equiv\frac{1}{2}\left[
\frac{\mbox{BR}(B_d^0\to\pi^- K^+)+
\mbox{BR}(\bar B_d^0\to\pi^+ K^-)}{\mbox{BR}(B_d^0\to\pi^0K^0)+
\mbox{BR}(\bar B_d^0\to\pi^0\bar K^0)}\right],
\end{equation}
i.e.\ to consider separately the charged and neutral $B\to\pi K$ 
modes \cite{BF-neutral1}. The experimental situation of these quantities
is summarized in Table~\ref{tab:BpiK-input1}. We observe that $R_{\rm c}$
went down, thanks to the larger value of $\mbox{BR}(B^\pm\to\pi^\pm K)$,
and that $R_{\rm n}$ moved marginally up. 

Furthermore, the decay $B^+\to\pi^0K^+$ offers a direct CP asymmetry,
\begin{equation}\label{ACP-Bpi0K-def}
{\cal A}_{\rm CP}^{\rm dir}(B^\pm\to\pi^0 K^\pm)\equiv
\frac{\mbox{BR}(B^+\to\pi^0K^+)-
\mbox{BR}(B^-\to\pi^0 K^-)}{\mbox{BR}(B^+\to\pi^0K^+)+
\mbox{BR}(B^-\to\pi^0K^-)}=-0.04\pm0.04,
\end{equation}
where we have also given the experimental average \cite{HFAG}. 
In the case of $B^0_d\to\pi^0K_{\rm S}$ decays, we have a final state 
with CP eigenvalue $-1$. Consequently, we may introduce a time-dependent 
rate asymmetry with the same structure as (\ref{rate-asym}), exhibiting 
direct {\it and} mixing-induced CP asymmetries. The most recent values 
for these observables obtained by the BaBar \cite{BaBar-BK0pi0}
and Belle \cite{Belle-BK0pi0} collaborations are consistent with each 
other, and correspond to the following new averages \cite{HFAG}:
\begin{eqnarray}
{\cal A}_{\rm CP}^{\rm dir}(B_d\to\pi^0 K_{\rm S})&=&+0.09\pm0.14
\label{Adir-Bdpi0KS}\\
{\cal A}_{\rm CP}^{\rm mix}(B_d\to\pi^0 K_{\rm S})&=&
-(0.34^{+0.27}_{-0.29}).
\end{eqnarray}

\subsubsection{Confrontation with Theory}
The SM amplitudes for the decays $B^0_d\to\pi^-K^+$ and 
$B^+\to\pi^+K^0$ were already given in (\ref{Bdpi-K+}) 
and (\ref{B+pi+K0}), respectively. In the case of the $B^+\to\pi^0K^+$
and $B^0_d\to\pi^0K^0$ modes, the decay amplitudes can be
written in the following form within the SM:
\begin{eqnarray}
\sqrt{2}A(B^+\to\pi^0K^+)&=&P'\left[1+\rho_{\rm c}e^{i\theta_{\rm c}}
e^{i\gamma}-\left(e^{i\gamma}-qe^{i\omega}\right)
r_{\rm c}e^{i\delta_{\rm c}}\right]\label{B+pi0K+-SM}\\
\sqrt{2}A(B^0_d\to\pi^0K^0)&=&-P'\left[1+\rho_{\rm n}e^{i\theta_{\rm n}}
e^{i\gamma}-qe^{i\omega}r_{\rm c}e^{i\delta_{\rm c}}\right].
\label{B0pi0K0-SM}
\end{eqnarray}
Here the parameter $q$, with the CP-conserving strong phase $\omega$, 
measures the importance of the EW penguins with respect to the 
tree-diagram-like topologies. In the SM, it can be determined with the 
help of the $SU(3)$ flavour symmetry of strong interactions \cite{NR}, 
yielding
\begin{equation}\label{q-SM}
q e^{i\omega}=0.69 \times\left[\frac{0.086}{|V_{ub}/V_{cb}|}\right];
\end{equation}
for a detailed discussion of the colour-suppressed EW penguin contributions,
which are neglected in (\ref{B+pi0K+-SM}) and (\ref{B0pi0K0-SM}), 
see Appendix~D of \cite{BFRS-BIG}. Moreover, we have 
\begin{equation}\label{rc-def}
r_{\rm c}e^{i\delta_{\rm c}}\equiv\left(\frac{\lambda^2R_b}{1-\lambda^2}
\right)\left[\frac{{\cal T}'+{\cal C}'}{{\cal P}_t'-{\cal P}_c'}\right],
\end{equation}
as well as
\begin{equation}\label{rho-n-def}
\rho_{\rm n}e^{i\theta_{\rm n}}\equiv\left(\frac{\lambda^2R_b}{1-\lambda^2}
\right)\left[\frac{{\cal C}'+({\cal P}_t'-{\cal P}_u')}{{\cal P}_t'-
{\cal P}_c'}\right],
\end{equation}
where the notation is analogous to the one introduced in 
Subsection~\ref{ssec:Bpipi-ampls}; the primes remind us again that we
have now turned to $\bar b\to \bar s$ modes. We observe that the hadronic
parameters in (\ref{r-def}), (\ref{rc-def}) and (\ref{rho-n-def}) satisfy the 
following relations:
\begin{equation}\label{r-rho-rel}
r_{\rm c}e^{i\delta_{\rm c}}=re^{i\delta}+\rho_{\rm n}e^{i\theta_{\rm n}}
\end{equation}
\begin{equation}\label{rho-n-det}
\rho_{\rm n}e^{i\theta_{\rm n}}=re^{i\delta}x'e^{i\Delta'},
\end{equation}
with 
\begin{equation}\label{x-BpiK}
x'e^{i\Delta'}\equiv \frac{{\cal C}'+{\cal P}_{tu}'}{{\cal T}'-{\cal P}_{tu}'}.
\end{equation}
The values of $(r,\delta)$ following from the $B\to\pi\pi$
data can be found in (\ref{r-del-det}). In analogy, we may use 
\begin{equation}\label{x-rel-SU3}
x'e^{i\Delta'}=\left[\frac{f_\pi F_{BK}(M_\pi^2;0^+)}{f_K F_{B\pi}
(M_K^2;0^+)}\right]xe^{i\Delta} 
\end{equation}
to determine $(x',\Delta')$ from their $B\to\pi\pi$ counterparts given
in Table~\ref{tab:Bpipi-hadr}. The factor
\begin{equation}
\left[\frac{f_\pi F_{BK}(M_\pi^2;0^+)}{f_K F_{B\pi}
(M_K^2;0^+)}\right]=1.05 \pm 0.18,
\end{equation}
where the numerical value refers to the light-cone sum-rule analysis of 
\cite{Ball}, describes factorizable $SU(3)$-breaking 
corrections. Since (\ref{r-det}) is not affected by $SU(3)$-breaking 
effects within factorization, such a factor is not present in the case of 
this relation \cite{RF-BsKK}. Finally, we obtain then, with the help of 
(\ref{rho-n-det}), the numerical values
\begin{equation}\label{rhon-thetan-det}
\rho_{\rm n}=0.12^{+0.05}_{-0.05}, \quad \theta_{\rm n}=
-\left(19.6_{-23.7}^{+17.6}\right)^\circ,
\end{equation}
and (\ref{r-rho-rel}) yields
\begin{equation}\label{rc-deltac-det}
r_{\rm c}=0.20^{+0.08}_{-0.06}, \quad \delta_{\rm c}=
\left(6.9^{+17.9}_{-13.4}\right)^\circ.
\end{equation}
Alternatively, $r_{\rm c}$ can be determined through the 
following well-known relation \cite{GRL}:
\begin{equation}\label{rc-alt-det}
r_{\rm c}=\sqrt{2}\left|\frac{V_{us}}{V_{ud}}\right|\frac{f_K}{f_\pi}
\sqrt{\frac{{\rm BR}(B^\pm\to \pi^\pm\pi^0)}{{\rm BR}(B^\pm\to \pi^\pm K^0)}}
=0.190 \pm 0.011,
\end{equation}
which relies on the $SU(3)$ flavour symmetry and the neglect of the
$\rho_{\rm c}$ term in (\ref{B+pi+K0}). 

Having the parameters in (\ref{rhon-thetan-det}) and (\ref{rc-deltac-det}) 
at hand, we may predict the values of $R_{\rm c}$ and $R_{\rm n}$ and of
the CP-violating observables of $B^\pm\to\pi^\pm K$,
$B_d\to\pi^0K_{\rm S}$ in the SM with the help of the formulae given in 
\cite{BFRS-BIG}. In the case of the charged modes, we obtain
\begin{equation}
\left.R_{\rm c}\right|_{\rm SM}=1.14 \pm 0.05 \quad (1.12)\,,
\end{equation}
\begin{equation}
\left. {\cal A}_{\rm CP}^{\rm dir}(B^\pm\to\pi^0 K^\pm) \right|_{\rm SM}=
0.04\,^{+0.11}_{-0.08} \quad (0.04)\,,
\end{equation}
where here (and in the following) the numbers with errors refer
to $\rho_{\rm c}=0$, and the central value for the case
$\rho_{\rm c}=0.05$, $\theta_{\rm c}=0^\circ$ is given in brackets.
We observe that the impact of $\rho_{\rm c}$ on $R_{\rm c}$ is 
significantly weaker than in the case of $R$ discussed in 
Subsection~\ref{ssec:rhoc-analysis}. This is due to the feature that
$\rho_{\rm c}$ enters $R$ already at ${\cal O}(\rho_{\rm c})$,
whereas it affects $R_{\rm c}$ through second order terms of
${\cal O}(\rho_{\rm c}^2)$ and ${\cal O}(\rho_{\rm c} r_{\rm c})$
\cite{NR}. On the other hand, $R_{\rm n}$ and the CP-violating
observables of $B_d\to \pi^0K_{\rm S}$ are {\it not} affected by
$\rho_{\rm c}$, so that we obtain the following SM predictions:
\begin{equation}
\left. R_{\rm n}\right|_{\rm SM}=1.11^{+0.04}_{-0.05}
\end{equation}
\begin{equation}
\left.{\cal A}_{\rm CP}^{\rm dir}(B_d\to\pi^0 K_{\rm S})\right|_{\rm SM}=
0.07\,^{+0.08}_{-0.11}, \quad
\left.{\cal A}_{\rm CP}^{\rm mix}(B_d\to\pi^0 K_{\rm S})\right|_{\rm SM}=
-(0.87\pm0.05).
\end{equation}

So far, we could accommodate all features of the $B$-factory data 
for the $B\to\pi\pi$ and $B\to\pi K$ modes in a satisfactory manner
in the SM. Now we observe that this is {\it not} the case for $R_{\rm n}$ 
and -- to a smaller extend -- for $R_{\rm c}$. As we have emphasized 
above, $R_{\rm n}$ does {\it not} depend on $\rho_{\rm c}$, so this 
parameter cannot be at the origin of this puzzle, in contrast to the 
case of $R$, and has, moreover, a minor impact on $R_{\rm c}$. 
Moreover, as we discussed in \cite{BFRS-BIG}, the colour-suppressed EW 
penguin topologies have {\it no} impact on $R_{\rm n}$ in our SM analysis, 
as they can be absorbed in a certain manner, but could affect $R_{\rm c}$. 
However, as we have seen in Subsection~\ref{ssec:rhoc-analysis}, the analysis
of $R$ disfavours anomalously large contributions of this kind, 
in contrast to the claims made in \cite{CKM-fitters}. Concerning 
$SU(3)$-breaking corrections, the agreement between (\ref{ACPdir-av}) 
and our SM prediction (\ref{ACP-pred}), the successful confrontation 
of (\ref{CP-rel}) with the data, and the emerging picture of the 
UT -- in perfect accordance with the SM -- discussed in 
Subsection~\ref{ssec:UT-det} do not indicate large corrections to 
(\ref{r-det}). Moreover, the agreement between (\ref{rc-deltac-det}) 
and (\ref{rc-alt-det}) indicates that the leading $SU(3)$-breaking 
effects are indeed described by the corresponding factors in 
(\ref{x-rel-SU3}) and (\ref{rc-alt-det}). So what could then be 
the origin of the puzzling pattern of the measured values of $R_{\rm n}$ 
and $R_{\rm c}$?

\begin{figure}
\begin{center}
\includegraphics[width=12cm]{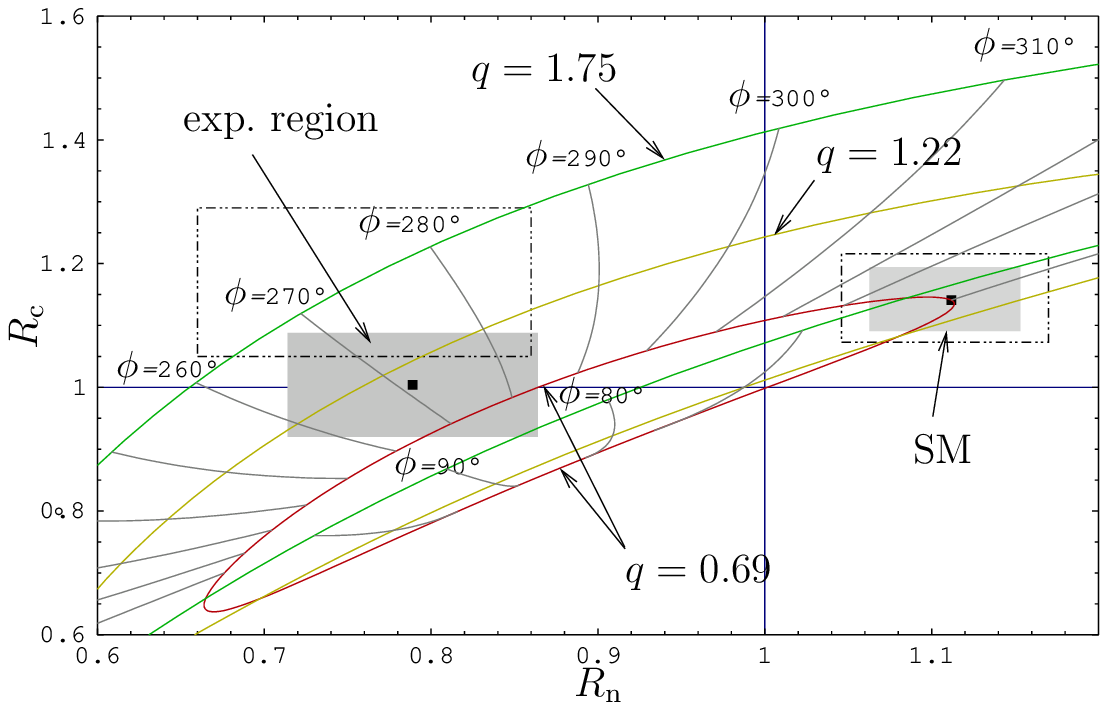}
\end{center}
\caption{The situation in the $R_{\rm n}$--$R_{\rm c}$ plane. We show contours
for values of $q=0.69$, $q=1.22$ and $q=1.75$, with
$\phi \in [0^\circ,360^\circ]$. 
The experimental ranges for $R_{\rm c}$ and $R_{\rm n}$ and those predicted 
in the SM are indicated in grey, the dashed lines serve as a reminder of
the corresponding ranges in \cite{BFRS-BIG}.\label{fig:Rn-Rc}}
\end{figure}

\subsubsection{NP in the EW Penguin Sector}\label{ssec:NP-EWP}
Since $R_{\rm n}$ and $R_{\rm c}$ are significantly affected by EW 
penguins, it is an attractive possibility to assume that NP enters 
through these topologies \cite{FM-NP,trojan}. In this case, the successful 
picture described above would not be disturbed. On the other hand, we may 
obtain full agreement between the theoretical values of $R_{\rm n}$
and $R_{\rm c}$ and the data. Following \cite{BFRS-PRL,BFRS-BIG}, we 
generalize the EW penguin parameter as
\begin{equation}
q \to q e^{i\phi},
\end{equation}
where $\phi$ is a CP-violating weak phase that vanishes in the SM,
i.e.\ arises from NP. We may then use the measured values of 
$R_{\rm c}$ and $R_{\rm n}$ to determine $q$ and $\phi$, with
the following results:
\begin{equation}\label{q-det}
q = 1.08\,^{+0.81}_{-0.73} \quad (1.23)\,,
\end{equation}
\begin{equation}\label{phi-det}
\phi = -(88.8^{+13.7}_{-19.0})^\circ \quad (-86.8^\circ)\,,
\end{equation}
where the numbers in brackets illustrate again the impact of
$\rho_{\rm c}=0.05$, $\theta_{\rm c}=0^\circ$ on the central values. 
Although these hadronic parameters are not at the origin of the
$B\to \pi K$ puzzle, as we have seen above, they have of course
some impact on the extracted values of $q$ and $\phi$, though
the are not changing the overall picture. 

It is useful to consider the $R_{\rm n}$--$R_{\rm c}$ plane, as 
we have done in Fig.~\ref{fig:Rn-Rc}. There we show contours corresponding 
to different values of $q$, and indicate the experimental and SM ranges. 
Following \cite{BFRS-BIG}, we choose the values of $q=0.69$, $1.22$ and 
$1.75$, where the latter reproduced the central values of $R_{\rm c}$ 
and $R_{\rm n}$ in our previous analysis \cite{BFRS-PRL,BFRS-BIG}. The 
central values for the SM prediction have hardly moved, while their
uncertainties have been reduced a bit. On the other hand, the central 
experimental values of $R_{\rm c}$ and $R_{\rm n}$ have moved in such a 
way that $q$ decreased, while the weak phase $\phi$ remains around 
$-90^{\circ}$. 

Moreover, we obtain the following CP asymmetries in our NP scenario:
\begin{equation}
{\cal A}_{\rm CP}^{\rm dir}(B^\pm\to\pi^0 K^\pm)= 
  0.10^{+0.25}_{-0.19} \quad (0.10)
\end{equation}
\begin{eqnarray}
{\cal A}_{\rm CP}^{\rm dir}(B_d\to\pi^0 K_{\rm S})=0.01\,^{+0.15}_{-0.18}, 
\quad 
{\cal A}_{\rm CP}^{\rm mix}(B_d\to\pi^0 K_{\rm S})=-(0.98\,^{+0.02}_{-0.04}).
\end{eqnarray}
Although the central value of our prediction for ${\cal A}_{\rm CP}^{\rm
dir}(B^\pm\to\pi^0 K^\pm)$ has a sign different from the corresponding 
experimental number, this cannot be considered as a problem because of
the large current uncertainties. Concerning the observables of the
$B_d\to\pi^0 K_{\rm S}$ channel, our prediction for the direct CP asymmetry 
is rather close to the experimental number, while the current 
experimental result for mixing-induced asymmetry is somewhat on
the lower side. However, the uncertainties of these very challenging 
measurements are still too large to draw conclusions. 

Instead of using the value of $\omega=0^\circ$ in our analysis,
which follows from the $SU(3)$ flavour symmetry, we could alternatively
determine this strong phase, together with $q$ and $\phi$, from
a combined analysis of $R_{\rm n}$, $R_{\rm c}$ and the direct
CP asymmetry of the $B^\pm\to\pi^0 K^\pm$ modes. In our previous 
analysis $\cite{BFRS-BIG}$, this led to small values of $\omega$
in perfect agreement with the picture following from the $SU(3)$
flavour symmetry. Using the most recent data, we obtain
\begin{equation}\label{omega-det}
\omega=-\left(20\,^{+43}_{-28}\right)^{\circ},\quad
q = 1.08\,^{+0.82}_{-0.67},\quad
\phi = -(88.2^{+14.0}_{-21.0})^\circ \,,
\end{equation}
where the values of $q$ and $\phi$ are practically unchanged from 
the numbers given in (\ref{q-det}) and (\ref{phi-det}), respectively. 
The updated value in (\ref{omega-det}) does still not favour
dramatic $SU(3)$-breaking effects.

Finally, we would like to comment briefly on the direct CP asymmetry
of the $B^\pm\to\pi^0K^\pm$ decays. It was argued in the recent
literature (see, for instance, \cite{ligeti}) that the discrepancy 
between the experimental values in (\ref{ACPdir-av}) and 
(\ref{ACP-Bpi0K-def}) was very puzzling. However, our analysis shows
nicely that this is actually not the case. In particular, we
have the following expression \cite{BFRS-BIG}:
\begin{equation}\label{ACP-Bpi0K-defa}
{\cal A}_{\rm CP}^{\rm dir}(B^\pm\to\pi^0 K^\pm)=
\frac{2}{R_{\rm c}}\left[r_{\rm c}\sin\delta_{\rm c}\sin\gamma
-qr_{\rm c}\left\{\sin(\delta_{\rm c}+\omega)\sin\phi+r_{\rm c}
\sin\omega\sin(\gamma-\phi)\right\}\right],
\end{equation}
where the $\rho_{\rm c}$ terms are neglected for simplicity. 
Consequently, the small value in (\ref{ACP-Bpi0K-def}) follows
simply from the small strong phase $\delta_{\rm c}$ in 
(\ref{rc-deltac-det}). On the other hand, the hadronic parameters
$r$ and $\delta$ governing ${\cal A}_{\rm CP}^{\rm dir}(B_d\to\pi^\mp K^\pm)$
take very different values, as we have seen in Subsection~\ref{ACP-dir-theo1}.
The difference between the CP-violating $B^\pm\to\pi^0 K^\pm$ and 
$B_d\to\pi^\mp K^\pm$ asymmetries can therefore be straightforwardly
explained through hadronic effects within the SM, i.e.\ does not
require NP. 

\boldmath
\section{Conclusions}\label{sec:concl}
\unboldmath
\setcounter{equation}{0}
In this paper we have confronted our strategy for describing
and correlating the $B\to\pi\pi$, $B\to\pi K$ decays and rare 
$K$ and $B$ decays with the new data on $B\to\pi\pi$, $B\to\pi K$ from 
BaBar and Belle. Within a simple NP scenario of enhanced 
CP-violating EW penguins considered by us, the NP contributions 
enter significantly only $B\to\pi K$ decays and rare $K$ and $B$ decays, 
while the $B\to\pi\pi$ system is practically unaffected by these 
contributions and can be described within the SM. Consequently the 
pattern of relations between various observables is in our strategy very
transparent in that

\begin{itemize}
\item   
The relations between $B\to\pi\pi$ and $B\to\pi K$ decays are strictly
connected with the {\it long-distance} physics allowing us to calculate the
hadronic parameters of the $B\to\pi K$ from the $B\to\pi \pi$ ones 
without the intervention of NP contributions that enter at much
shorter scales.
\item
The relations between $B\to\pi K$ decays and rare $K$ and $B$
decays are strictly connected with the {\it short-distance} physics
allowing us to predict several spectacular
departures from the SM expectations for rare $K$ and $B$ decays
from the corresponding significant departures
from the SM observed in the  $B\to\pi K$ data.
\end{itemize}

The main messages from this new analysis are as follows:
\begin{itemize}
\item
The present data for those observables in the $B\to\pi\pi$ and
$B\to\pi K$  systems that are essentially unaffected by NP in the
EW penguins are not only in  accordance with our approach,
but a number of predictions made by us in \cite{BFRS-PRL,BFRS-BIG} 
have been confirmed by the new data within theoretical and
experimental uncertainties. This is in particular the case
of the direct CP asymmetry in $B_d\to \pi^\mp K^\pm$ but also in
$B_d\to\pi^0\pi^0$. For convenience of the reader, we collect all
CP-violating quantities involved in our analysis in Table \ref{asymtab} and 
show the comparison of the experimental values with our predictions.
\item The observed decrease of the ratio $R$ below our
expectations in \cite{BFRS-PRL,BFRS-BIG}  can be partially attributed to
certain hadronic effects, represented by the non-vanishing value of
$\rho_{\rm c}$, that could be tested in $B^\pm\to K^\pm K$ 
decays once these are experimentally better known. 
In particular the sign of $\rho_{\rm c}$, which is more solid than its 
magnitude, points towards the decrease of $R$ relative to our previous
estimate. However, our present
understanding of these effects allows us to expect that future more
accurate measurements will find $R$ higher than its present
central value.
\item The decrease in the difference $R_{\rm c}-R_{\rm n}$ observed in the
recent data of BaBar has been predicted by us on the basis of
branching ratios for rare decays \cite{BFRS-PRL,BFRS-BIG}. 
This is explicitly seen in Table 2 of \cite{BFRS-BIG}.
 In this manner
the overall description of $B\to \pi \pi$, $B\to \pi K$ and rare decays 
within our
approach has improved with respect to our previous analysis.
\item The picture of rare decays presented by us in \cite{BFRS-PRL,BFRS-BIG}
remains unchanged, since the values of $q$ and $\phi$ obtained are
still slightly above the bound from $b \to s l^+ l^-$ used in
\cite{BFRS-BIG}. In particular, the spectacular enhancement of
$K_{\rm L}\to\pi^0\nu\bar{\nu}$ as well as the enhancement of several 
other rare decays remain. Further implications for rare decays in this
scenario can be found in {\cite{RaiChoudhury:2004pw, Isidori:2004rb}}
\item Last but certainly not least the obtained value of $\gamma$ and
the UT are in full agreement with the usual CKM fits.
\end{itemize}
Finally, we would like to comment on analyses using only the $B \to \pi K$ 
data. It has been claimed in
\cite{CKM-fitters,CL,ILPL} that the puzzle concerning the $B \to \pi
K$ system is significantly reduced or not even present. 
We would like to emphasize that a study of the $B\to\pi K$ decays
alone is not very much constrained and consequently has a rather low
resolution in search for NP effects. Such an analysis is moreover not
satisfactory as it ignores the information 
on long distance dynamics that we have already from other non-leptonic 
decays, in particular from $B\to\pi\pi$ decays that are connected with the
$B\to\pi K$ system through $SU(3)$ flavour symmetry. One should also
not forget 
that, for a confrontation with the SM, the use of the $SU(3)$ flavour 
symmetry, which allows us to determine the EW penguin parameters $q$ 
and $\omega$ through (\ref{q-SM}), cannot be avoided.
Consequently one may ask why the $SU(3)$ flavour symmetry should be used 
to find $q$ and not for getting the full input from $B\to\pi\pi$ and in the
future from $B\to KK$ decays. 

As demonstrated in {\cite{BFRS-PRL, BFRS-BIG}} and here, the use of
the full information from the $B \to \pi \pi$ and $B \to \pi K$ systems
allows us to uncover possible signals of NP effects in $B \to \pi K$
decays that in turn change significantly the SM pattern of rare decay
branching ratios. In this respect, we disagree with a statement made
in \cite{ligeti} that ``the data seem to disfavour NP
explanations, according to which NP primarily modifies electroweak
penguin contributions''.

\begin{table}[hbt]
\vspace{0.4cm}
\begin{center}
\begin{tabular}{|c||c|c|}
\hline
  Quantity &   Our Prediction &  Experiment
 \\ \hline
  ${\cal A}_{\rm CP}^{\rm dir}(B_d\!\to\!\pi^0 \pi^0)$ 
 &   $-0.28^{+0.37}_{-0.21}$ \rule{0em}{1.05em} & $-0.28 \pm 0.39$ \\\hline
  ${\cal A}_{\rm CP}^{\rm mix}(B_d\!\to\!\pi^0 \pi^0)$ 
 &  $-0.63^{+0.45}_{-0.41}$ \rule{0em}{1.05em} &
 $-0.48_{-0.40}^{+0.48}$ \\\hline
${\cal A}_{\rm CP}^{\rm dir}(B_d\!\to\!\pi^{\mp} K^{\pm}) $ 
&   $0.127^{+0.102}_{-0.066}$ \rule{0em}{1.05em} &  $0.113\pm0.019$ \\
\hline
${\cal A}_{\rm CP}^{\rm dir}(B^\pm\!\to\!\pi^0 K^\pm) $ 
&   $0.10^{+0.25}_{-0.19}$ \rule{0em}{1.05em} &  $-0.04 \pm 0.04$ \\ \hline
  ${\cal A}_{\rm CP}^{\rm dir}(B_d\!\to\!\pi^0 K_{\rm S})$ 
&   $0.01^{+0.15}_{-0.18}$ \rule{0em}{1.05em} &
$0.09 \pm 0.14$ \\ \hline 
  ${\cal A}_{\rm CP}^{\rm mix}(B_d\!\to\!\pi^0 K_{\rm S})$ 
&  $-0.98^{+0.04}_{-0.02} $ \rule{0em}{1.05em} &  $-0.34^{+0.29}_{-0.27}$
\\ \hline
\end{tabular}
\caption{Compilation of predictions for all CP-violating asymmetries
  in the $B \to \pi \pi$ and $B \to \pi K$ systems. We omit the quantities used as input.} 
\label{asymtab}
\end{center}
\end{table}

It will be exciting to follow the experimental progress on
$B\to\pi\pi$ and $B\to\pi K$ decays and the corresponding efforts in rare
decays. In particular new messages from BaBar and Belle that 
the present central values of $R_{\rm c}$ and $R_{\rm n}$ have been confirmed 
at a high confidence level, a slight increase of $R$ and a message 
from KEK {\cite{KEK}} in the next two
years that the decay $K_{\rm L}\to\pi^0\nu\bar\nu$ has been observed would 
give a strong support to the NP scenario considered here.

\vspace*{0.5truecm}


\noindent
{\bf Acknowledgments}\\
\noindent
The work presented here was supported in part by the German 
Bundesministerium f\"ur
Bildung und Forschung under the contract 05HT4WOA/3 and the 
DFG Project Bu.\ 706/1-2.


%
%
%
\begin{appendix}
\boldmath
\section{Predictions for the $B_s\to K^+K^-$ Observables}\label{app:BsKK}
\unboldmath
\setcounter{equation}{0}
The decay $B_s\to K^+K^-$ is related to $B_d\to\pi^+\pi^-$ through
the interchange of all down and strange quarks, i.e.\ through the
$U$-spin flavour symmetry of strong interactions. Consequently,
this symmetry allows us to determine the hadronic $B_s\to K^+K^-$ 
parameters $(d',\theta')$ through the values of their $B_d\to\pi^+\pi^-$
counterparts $(d,\theta)$ given in Table~\ref{tab:Bpipi-hadr}. Using
then the range of $\gamma$ in (\ref{gamma-range}), and the SM value
$\phi_s=-2^\circ$ for the $B^0_s$--$\bar B^0_s$ mixing phase, 
we arrive at the following SM predictions, updating those given
in \cite{BFRS-BIG}:
\begin{eqnarray}
{\cal A}_{\rm CP}^{\rm dir}(B_s\to K^+K^-)&=&
0.13^{+0.10}_{-0.07}\label{ACP-dir-BsKK}\\
{\cal A}_{\rm CP}^{\rm mix}(B_s\to K^+K^-)&=&
-0.18 \pm 0.05.\label{ACP-mix-BsKK}
\end{eqnarray}
Concerning the CP-averaged $B_s\to K^+K^-$ branching ratio, which
is of more immediate experimental interest, we have to take a certain
$SU(3)$-breaking factor into account that has recently been calculated
through QCD sum rules \cite{KMM}. Following \cite{BFRS-BIG}, we obtain 
the updated value
\begin{equation}\label{BR-BsKK}
\mbox{BR}(B_s\to K^+K^-)=(38^{+32}_{-23})\times 10^{-6}
\end{equation}
from the $B\to\pi\pi$ data. Alternatively, we may calculate
$\mbox{BR}(B_s\to K^+K^-)$ with the help of the CP-averaged
$B_d\to \pi^\mp K^\pm$ branching ratio, which requires, however, 
the additional assumption that penguin annihilation and exchange
topologies play a minor r\^ole (see item ii in 
Subsection~\ref{ssec:BpiK-pre}) . Following this avenue yields
\begin{equation}\label{BR-BsKK-alt}
\mbox{BR}(B_s\to K^+K^-)=(35 \pm 7)\times 10^{-6},
\end{equation}
in nice agreement with (\ref{BR-BsKK}). Let us note that the 
difference between (\ref{ACP-dir-BsKK})--(\ref{BR-BsKK}) and the 
corresponding numbers in \cite{BFRS-BIG} is very small, whereas
(\ref{BR-BsKK-alt}) did not change at all. 

The CDF Collaboration has recently reported the first measurements
of the CP-averaged $B_s\to K^+K^-$ branching ratio \cite{CDF}, 
corresponding to the preliminary result 
\begin{equation}
\mbox{BR}(B_s\to K^+K^-)=(34.3 \pm 5.5 \pm 5.2)\times 10^{-6}.
\end{equation}
The agreement with our theoretical SM predictions is very impressive, 
giving further support to our strategy. We look forward to better data
and hope that also first measurements of the CP-violating $B_s\to K^+K^-$ 
observables will be available in the near future. Here 
${\cal A}_{\rm CP}^{\rm mix}(B_s\to K^+K^-)$ would be particularly
exciting, since this asymmetry may well be affected by NP contributions 
to $B^0_s$--$\bar B^0_s$ mixing, which would manifest themselves then
as a discrepancy to (\ref{ACP-mix-BsKK}). By the time this measurement
will be available, the uncertainty of the SM prediction given there 
should be further reduced thanks to better $B_d\to\pi^+\pi^-$ 
input data.

\end{appendix}
%
%
%


%
%
%

%
%
%
\end{document}